\newcommand{\be}{\begin{equation}}
\newcommand{\ee}{\end{equation}}
\newcommand{\bea}{\begin{eqnarray}}
\newcommand{\eea}{\end{eqnarray}}
\newcommand{\beann}{\begin{eqnarray*}}
\newcommand{\eeann}{\end{eqnarray*}}
\newcommand{\beasn}{\begin{sneqnarray}}
\newcommand{\eeasn}{\end{sneqnarray}}
\newcommand{\ba}{\begin{array}}
\newcommand{\ea}{\end{array}}

\newcommand{\Appendix}[1]%
    {\renewcommand{\thesection}{Appendix~\Alph{section}:}%
     \section{#1}}%
\catcode`@=11
\long\def\@makecaption#1#2{
   \vskip 10pt
   \setbox\@tempboxa\hbox{{\small\bf #1.} \ {\small #2}}
   \ifdim \wd\@tempboxa >\hsize       % IF longer than one line:
   {\small\bf #1.} \ {\small #2}\par  % THEN set as ordinary paragraph.
   \else                              %   ELSE  center.
        \hbox to\hsize{\hfil\box\@tempboxa\hfil}
   \fi}
\catcode`@=12

\catcode`@=11
\def\secteqno{\@addtoreset{equation}{section}%
\def\theequation{\thesection.\arabic{equation}}}
\def\endsecteqno{\def\theequation{\@ifundefined{chapter}%
{\arabic{equation}}{\thechapter.\arabic{equation}}}}
\newcounter{subequation}
\def\thesubequation{\alph{subequation}}
\def\sneqnarray{\stepcounter{equation}\let\@currentlabel=\theequation
\setcounter{subequation}{1}
\def\@eqnnum{{\rm (\theequation\thesubequation)}}
\global\@eqcnt\z@\tabskip\@centering\let\\=\@eqncr\let\@@eqncr=\@@sneqncr
$$\halign to \displaywidth\bgroup\@eqnsel\hskip\@centering
 $\displaystyle\tabskip\z@{##}$&\global\@eqcnt\@ne
 \hskip 2\arraycolsep \hfil${##}$\hfil
 &\global\@eqcnt\tw@ \hskip 2\arraycolsep
$\displaystyle\tabskip\z@{##}$\hfil
  \tabskip\@centering&\llap{##}\tabskip\z@\cr}
\def\endsneqnarray{\@@sneqncr\egroup $$\global\@ignoretrue}
\def\@@sneqncr{\let\@tempa\relax
   \ifcase\@eqcnt \def\@tempa{& & &}\or \def\@tempa{& &}
   \else \def\@tempa{&}\fi
     \@tempa \if@eqnsw\@eqnnum\stepcounter{subequation}\fi
     \global\@eqnswtrue\global\@eqcnt\z@\cr}
\def\nobiblabels{\def\@lbibitem[##1]##2{\@bibitem{##2}}}
\catcode`@=12

\def\a{\alpha}

  \def\L{\Lambda}  
    
\documentclass[12pt]{article}
\usepackage[german,english]{babel}
\usepackage{amssymb}
\usepackage{amsmath}                                                         
\usepackage{epsfig,boxedminipage}

\secteqno
\renewcommand{\thesection}{\arabic{section}.}
\renewcommand{\thesubsection}{\arabic{section}.\arabic{subsection}}
\renewcommand{\theequation}{\thesection  \arabic{equation}}

\begin{document}

%%%%%%%%%%%%%%%%%%%%%%%%%%%%%%%%%%% TITLE %%%%%%%%%%%%%%%%%%%%%%%%%%%%%%%%%%%%%%

\title{{\bf Renormalizing the Lippmann-Schwinger equation\\ for the one pion exchange potential}} 
\author{ {\Large {\sl Dolors Eiras}} {\sl and} {\Large {\sl Joan Soto}}\\
        \small{\it{Departament d'Estructura i Constituents de la Mat\`eria 
                   and IFAE}}\\
        \small{\it{Universitat de Barcelona}}\\
        \small{\it{Diagonal, 647, E-08028 Barcelona, Catalonia, Spain.}}\\  \\
        {\it e-mails:} \small{dolors@ecm.ub.es, soto@ecm.ub.es} }
\date{\today}

\maketitle

\thispagestyle{empty}

%\enlargethispage{\baselinestretch}

\begin{abstract}
We address the question whether the cut-off dependence, which has to be introduced in order to properly define the Lippmann-Schwinger equation for the one pion exchange potential plus local ($\delta$-function) potentials, can be removed (up to inverse powers of it) 
by a suitable tuning of the various (bare) coupling constants. We prove that this is indeed so both for the spin singlet and for the spin triplet channels. 
However, the latter requires such a strong cut-off dependence of the coupling constant associated to the non-local term which breaks orbital angular momentum conservation, that the renormalized amplitude lacks from partial wave mixing. We argue that this 
is an indication that this term must be treated perturbatively.

\end{abstract}

%\bigskip
PACS: 03.65.Nk, 11.10.Gh, 13.75.Cs, 21.30.Fe, 21.45.+v .

\vfill
\vbox{
\hfill{}\null\par
\hfill{UB-ECM-PF 01/06}\null\par}

\newpage

%%%%%%%%%%%%%%%%%%%%%%%%%%%%%%%%%%%%%%%%%%%%%%%%%%%%%%%%%%%%%%%%%%%%%%%%%%%%%%%%
%%%%%%%%%%%%%%%%%%%%%%%%%%%%%% INTRODUCTION %%%%%%%%%%%%%%%%%%%%%%%%%%%%%%%%%%%%

\section{Introduction}
\indent

Since the original suggestion by Weinberg \cite{Weinberg} that the nuclear forces could be understood within the framework of effective field theories (EFT) there has been an increasing interest in the subject (see \cite{reviews} for recent reviews). A key ingredient of the EFT formalism is that the cut-off dependence which is introduced in order to smooth out ultraviolet (UV) singularities can be absorbed by suitable counterterms, and hence any dependence on physical scales much higher than the ones of the problem at hand can be encoded in a few (unknown) constants. In order to achieve this in a systematic manner counting rules are also necessary.

Weinberg's suggestion consisted of two steps. The first one was calculating the nucleon-nucleon (NN) potentials order by order in Chiral Perturbation Theory
($\chi$PT) 
from the Heavy Baryon Chiral Lagrangian (HB$\chi$L) \cite{JM}. The second one introducing the potentials thus obtained in a Lippmann-Schwinger (LS) equation. There is no doubt that the first step can be carried out within an EFT framework: the renormalized NN potentials are known
at leading, NL and NNL order \cite{VK,Meissner}, with even higher order corrections \cite{Kaiser} and isospin breaking terms \cite{Walzl} taken care of. The second step however is delicate. The potentials obtained in the first step are increasingly singular at 
short distances as we rise the order of $\chi$PT they are calculated. Hence the introduction of a regulator in the LS equation is compulsory. Since, even with the leading order (LO) potential, the LS equation can only be solved numerically, it is not clear that the scattering amplitude thus obtained is cut-off independent.
This is so even for the successful fits \cite{VK,Meissner} to different partial amplitudes, where the cut-off is regarded as a variational parameter close to the last scale integrated out. We present here a proof that this cut-off can  be removed from the LO (in the $\chi$PT counting) NN interaction if we tune properly the coupling constants of the potential. However, for this to be so 
we also have to tune the coupling constant of a non-local potential in the triplet channel. Even then, the only solution we find turns out to be physically unacceptable. Nevertheless, the insight on scaling so gained enables us to put forward a new proposal of counting rules where, coming back to standard procedures, divergences are fully absorbed by local counterterms.  

As EFTs have been mainly used in a perturbative framework, it is far from obvious how the two main features of those, namely renormalizability and counting rules, must be implemented in a non-perturbative one. Although in this work we shall primarily address the question of renormalizability, we would like to
start by making a 
 remark on counting rules, which emanates from previous experience on EFT in non-perturbative systems. It was pointed out in ref. \cite{Mont} that calculating the potential in a non-relativistic system can be understood as the integration of certain degrees of freedom, which can be implemented as a matching calculation between two EFTs. 
In our case the higher energy EFT is the HB$\chi$L for the two nucleon sector, which is a local theory with pions and non-relativistic nucleons as explicit degrees of freedom. This EFT has an energy ($E$) cut-off ($\Lambda_E$) such that $E\sim m_{\pi} << \Lambda_E << M\sim 4\pi f_{\pi}$ and a momentum ($p$) cut-off ($\Lambda_p$) such that $p << \Lambda_p << M\sim 4\pi f_{\pi}$ ($m_{\pi}$ and $M$ stand for the pion and nucleon mass respectively, and $4\pi f_{\pi}$ for the scale of non-Goldstone boson QCD states). Its lagrangian can be organized according to the chiral counting since chiral symmetry (and its breaking) are explicit. The lower energy EFT has an energy cut-off ($\Lambda_E^{\prime}$) such that $E<< \Lambda_E^{\prime}<< m_{\pi} $ and a momentum cut-off such that $p \lesssim m_{\pi} << \Lambda_p^{\prime}<< M $. It consists of non-relativistic nucleons interacting through a (non-local) potential.
The potential plays the role of a matching coefficient. As such, the potential encodes information on the higher energy EFT and it can be calculated {\it independently} on how the calculation of the lower energy EFT is organized, namely {\it independently} on what the counting in the low energy EFT is. Hence, on the one hand, the potential can be calculated order by order in $\chi$PT. On the other hand, chiral symmetry is not explicit anymore in the lower energy EFT (no pion fields exist) and consequently the chiral counting is not the natural way to organize the calculation anymore.

An interesting example of a related situation is the pionium system
(see \cite{Bern3} for a recent account), which has been studied using a series of EFTs \cite{Pionium}. The higher energy EFT is the 
Chiral Lagrangian coupled to electromagnetism  and the lower energy one a quantum mechanical Hamiltonian with the Coulomb potential and local interactions.
The matching between the two EFTs can be carried out perturbatively in $\chi$PT and $\alpha$, but the calculations in the lower EFT are carried out keeping
the Coulomb potential non-perturbatively (otherwise no bound state exists) and, furthermore, one does not need to specify to which order of $\chi$PT the local
potentials have been calculated.

Following that spirit, the main question for the NN system is what 
should be treated as
the LO potential in the low energy calculations. In the 
(higher energy) $\chi$PT counting the LO potential consists of the one pion exchange term (OPE) plus two local ($\delta$-function) terms. This assumes that the  natural 
scale of the two local terms is of the order of the last scale integrated out ($\sim M$).
If the NN system was in a perturbative regime the scale of these two local terms would 
provide the scale of the scattering lengths. Since the experimental scattering lengths are much larger than the ones 
predicted in this way, we can foresee at least two possibilities. The first one is that
an unsuspected behavior of QCD at energies $\sim \Lambda_{QCD}$ produces unnaturally large values for the local potentials. Then one may consider these local terms as the (low energy) LO potential and treat the OPE (and higher orders) perturbatively \cite{KSW}. This approach has been worked out  at ${\rm N^2LO}$ \cite{Stewart} showing slow-convergence in the $^1S_0$ channel and no convergence at all in the $^3S_1$-$^3D_1$ channel. 
The second possibility is that the local terms do have natural sizes but the low-energy dynamics is responsible for the large scattering lengths. In this case there is no reason
to treat the OPE perturbatively and a fully non-perturbative evaluation of the LS equation with LO potential (in the $\chi$PT counting) is required \cite{VK,Meissner}. We shall stick to this second possibility for most of the paper, although eventually a third
possibility, which is half way, will emerge as the most reasonable one (to us).

Before going on, let us briefly discuss some previous work on the renormalization of the LS equation. 
The case of a local (and hence separable) potential,
namely consisting of delta functions and its derivatives, has received plenty of attention\cite{Phillips,Beane,Steele}. This was expected to mimic the very low-energy ($p<<m_{\pi}$)
behavior 
of NN scattering. The regularization of this pure local EFT was a matter of debate some time ago: a cut-off regularization showed a systematic order by order improvement in the phase shift fit whereas dimensional regularization (DR) with MS scheme was extremely sensible to the large scattering length and shallow (nearly)-bound state, which translated into a poor radius of convergence. The shortcomings of DR with MS  were cured using the 
PDS scheme \cite{KSW} (see also \cite{StewartMehen}). The final outcome appears to be equivalent to the well-known Effective Range Expansion \cite{Kolck}. The next step in difficulty is renormalizing the LO potential in the $^1S_0$ channel, which contains a non-separable piece from the OPE. It was first carried out in \cite{Kaplan}, and reproduced by several
authors (see \cite{Bedaque}, for a recent report). We shall re-obtain these results in section 3.
Finally, as for renormalization in the $^3S_1$-$^3D_1$ channel, the available literature is, on the contrary, somewhat scarce \cite{Bedaque,Frederico} and the results are, to our understanding, not fully satisfactory (see sec. 6). 

The main difference of our approach with respect to the previous ones is that, in addition to the bare constants associated to local terms in the potential, we will also allow, but only in a initial stage, the bare constants of the non-local potentials to have non-trivial flows \footnote{In fact, this turns out to be the usual approach in theoretical works on renormalization of singular potentials (see for instance \cite{LaPlata}). Besides, there are known examples in a non-relativistic EFT of QCD (pNRQCD) where the renormalization of non-local potentials is required in order to absorb certain divergences \cite{LL}, the most spectacular of which being the renormalization of the static potential \cite{short}.}. This is less restrictive than the standard assumption that only local terms should renormalize the LS equation \cite{Lepage} for NN systems, which, in any case, is contained in it. This will permit us, not only to make meaningful comparisons with related work, but also to draw restrictions on the power counting. Having examined which conditions on the coupling constants are required in order to renormalize the LS equation, and the eventual consequences this has on our observables, we will be able to glean which terms of the potential can be included at LO, and which ones must be treated as perturbations, in the relevant case where only the coupling constants of local terms are allowed to flow and non-local potentials are fixed at the HB$\chi$PT values. 

So, once the (low energy) LO potential has been identified, we suspect that, in
order to be renormalizable, a higher order calculation should be organized as follows. The LS equation must be solved and renormalized treating the LO potential, as well as its couplings, non-perturbatively, but the NLO potentials and higher perturbatively.    

Therefore, the first step in this program is to identify a (low energy) LO potential 
 and to prove that it is renormalizable.
We start by the na\"\i ve choice, namely the LO potential in the $\chi$PT counting. For the spin singlet channel the LS equation is indeed renormalizable. However, for the spin triplet channels, we find that the LO potential in the $\chi$PT counting is only renormalizable if a certain coupling constant of a non-local potential has a non-trivial flow. (Or, in other words, if only the coupling constants of the local potentials are allowed to flow it is non-renormalizable.)
Even in this case, the physical outcome is not satisfactory: the partial wave mixing is washed out of the renormalized amplitude. We conclude that the (low energy) LO potential is not the full 
LO potential in the $\chi$PT
counting for the triplet channels. We identify a (low energy) LO potential, which is renormalizable, and prove that, if we treat the difference as a perturbation, it is also renormalizable at first order.

We distribute the paper as follows. In section 2 we introduce a convenient basis for the NN wave functions and our notations. A brief note at the end of this section serves to close all what refers to the isosinglet-singlet channel. In section 3 we prove that the isovector-singlet channel is renormalizable and provide explicit expressions for the cut-off dependence of the bare parameters both for a hard cut-off and for dimensional regularization. In section 4 we prove that the isosinglet-triplet channel is also renormalizable, but requires a strong cut-off dependence of the coupling constant of the (non-local) term in the potential which, in turn, prevents the renormalized amplitude from partial wave mixing. We interprete this result as an indication that this term must be treated perturbatively and prove that, if so, the first order in perturbation theory is finite. 
After briefly discussing in section 5 the isovector-spin vector channel, 
section 6 is devoted to a discussion. 

%%%%%%%%%%%%%%%%%%%%%%%%%%% DECOMPOSITION  %%%%%%%%%%%%%%%%%%%%%%%%%%%%%%%

\section{A convenient decomposition} 
\indent

We start from the LO NN potential given for instance in ref. \cite{Meissner}:
\begin{eqnarray}
V({\bf k}, {\bf k^{\prime}}) &=& 
-\left ( \frac{g_A}{2f_{\pi}} \right )^2 {\bf \tau}_1 \cdot {\bf \tau}_2  \frac{{\bf \sigma}_1 \cdot ({\bf k}-{\bf k^{\prime}}) \, {\bf \sigma}_2 \cdot ({\bf k}-{\bf k^{\prime}})}{({\bf k}-{\bf k^{\prime}})^2+m_{\pi}^2} + C_S + C_T \, {\bf \sigma}_1 \cdot {\bf \sigma}_2 \, ,
\label{(2.1)}
\end{eqnarray}
where $f_{\pi}$ is the pion decay constant ($\sim$ 93 MeV).

This potential acts on a wave function $\Psi^{ab}_{\alpha\beta}({\bf k}, {\bf k}^{\prime})$, where $a,b$ and $\alpha ,\beta$ are nucleon 
isospin and spin indices respectively. This wave function can be decomposed into irreducible representations of spin and isospin as 
follows:
\begin{eqnarray}
\Psi^{ab}_{\alpha\beta}({\bf k}) &=& \frac{1}{2} \left [ (\tau_2)^{ab} (\sigma_2)_{\alpha \beta} \, \psi_{SS}({\bf k}) + (\tau_2)^{ab} ({\bf \sigma}_{k^{\prime}} \sigma_2)_{\alpha \beta} \, \psi_{SV}^{k^{\prime}}({\bf k}) + \right. \nonumber \\
&&\left. +  ({\bf \tau}_{k} \tau_2)^{ab} (\sigma_2)_{\alpha \beta} \, \psi_{VS}^{k}({\bf k}) + ({\bf \tau}_{k} \tau_2)^{ab}({\bf \sigma}_{k'} \sigma_2)_{\alpha \beta} \,  \psi_{VV}^{kk'} ({\bf k}) \right ] \, .
\label{(2.1.1)}
\end{eqnarray}

The potential (\ref{(2.1)}) reduces for each isospin-spin channel to:
\begin{eqnarray}
V_{SS}({\bf k}, {\bf k^{\prime}}) &=&  -3 \, \left ( \frac{g_A}{2f_{\pi}} \right )^2 \frac{ ({\bf k}-{\bf k^{\prime}})^2}{({\bf k}-{\bf k^{\prime}})^2+m_{\pi}^2} + C_S - 3C_T \, , \nonumber \\
V_{SV}^{i'j'}({\bf k}, {\bf k^{\prime}}) &=&3 \,  \left ( \frac{g_A}{2f_{\pi}} \right )^2  \frac{({\bf k}-{\bf k^{\prime}})^2 \delta^{i'j'} -2({\bf k}-{\bf k^{\prime}})^{i'}({\bf k}-{\bf k^{\prime}})^{j'}}{({\bf k}-{\bf k^{\prime}})^2+m_{\pi}^2} + \left ( C_S +C_T \right ) \, \delta^{i'j'} \, , \nonumber \\
V_{VS}^{ij}({\bf k}, {\bf k^{\prime}}) &=& \left ( \frac{g_A}{2f_{\pi}} \right )^2 \frac{({\bf k}-{\bf k^{\prime}})^2 \delta^{ij}}{({\bf k}-{\bf k^{\prime}})^2+m_{\pi}^2} + \left ( C_S - 3C_T \right ) \, \delta^{ij} \, , \nonumber \\
V_{VV}^{ij, i'j'}({\bf k}, {\bf k^{\prime}}) &=& -\left ( \frac{g_A}{2f_{\pi}} \right )^2 \delta^{ij} \frac{({\bf k}-{\bf k^{\prime}})^2 \delta^{i'j'} -2({\bf k}-{\bf k^{\prime}})^{i'}({\bf k}-{\bf k^{\prime}})^{j'}}{({\bf k}-{\bf k^{\prime}})^2+m_{\pi}^2} + \left ( C_S +C_T \right ) \delta^{ij} \, \delta^{i'j'}  . \quad \quad
\label{(2.2)}
\end{eqnarray}

We still have to implement Fermi symmetry. This implies that the irreducible wave functions (\ref{(2.1.1)}) must fulfill (isospin and spin indices will be omitted for the rest of this section):
\begin{eqnarray}
\psi_{SS}({\bf k}) &=& -\psi_{SS}(-{\bf k}) \, , \nonumber \\
\psi_{SV}({\bf k}) &=& \psi_{SV}(-{\bf k}) \, , \nonumber \\
\psi_{VS}({\bf k}) &=& \psi_{VS}(-{\bf k}) \, , \nonumber \\
\psi_{VV}({\bf k}) &=& -\psi_{VV}(-{\bf k}) \, ,
\end{eqnarray}
which is implemented in the LS equation if we choose:
\begin{eqnarray}
T_{SS}({\bf k},{\bf k^{\prime}};E) = \frac{1}{2} \left ( V_{SS}({\bf k},{\bf k^{\prime}})-V_{SS}(-{\bf k},{\bf k^{\prime}})\right ) + \qquad \qquad \qquad \qquad \qquad  \nonumber \\
 +\frac{1}{2} \int^{\Lambda} \frac{d^3 k^{\prime \prime}}{(2\pi )^3} \, \left ( V_{SS}({\bf k},{\bf k^{\prime \prime}})-V_{SS}(-{\bf k}, {\bf k^{\prime \prime}})\right ) \frac{1}{E-\frac{{\bf k^{\prime \prime}}^2}{M}+i\eta}  T_{SS}({\bf k^{\prime \prime}},{\bf k^{\prime}};E) \qquad  (SS \longleftrightarrow VV) \, , \nonumber \\
T_{SV}({\bf k},{\bf k^{\prime}};E) =\frac{1}{2} \left ( V_{SV}({\bf k},{\bf k^{\prime}})+V_{SV}(-{\bf k},{\bf k^{\prime}})\right ) + \qquad \qquad \qquad \qquad \qquad \nonumber \\
+ \frac{1}{2}\int^{\Lambda} \frac{d^3 k^{\prime \prime}}{(2\pi )^3} \, \left ( V_{SV}({\bf k},{\bf k^{\prime \prime}})  +V_{SV}(-{\bf k},{\bf k^{\prime \prime}})\right ) \frac{1}{E-\frac{{\bf k^{\prime \prime}}^2}{M}+i\eta} T_{SV}({\bf k^{\prime \prime}},{\bf k^{\prime}};E) \nonumber \qquad (SV \longleftrightarrow VS) \, ,\\
\label{(2.5)}
\end{eqnarray}

It is the advantage of the above decomposition that we will not need to specify which (coupled) partial waves we are analyzing. 

If the LS equation for the potentials (\ref{(2.2)}) was well defined, using (\ref{(2.5)}) would be equivalent to solving the LS equation:
\begin{eqnarray}
{\widehat{T}}_{xy} ({\bf k}, {\bf k^{\prime}}; E) = V_{xy} ({\bf k},{\bf k^{\prime}})+ \int^{\Lambda} \, \frac{d^{3}k^{\prime \prime}}{(2\pi)^{3}} \, V_{xy} ({\bf k}, {\bf k^{\prime \prime}}) \, \frac{1}{E-\frac{{\bf k}^{\prime \prime 2}}{M}+i\eta} \, {\widehat{T}}_{xy}({\bf k^{\prime \prime}}, {\bf k^{\prime}}; E) \, ,
\label{(2.6)}
\end{eqnarray}
($x,y$=$S,V$), namely ignoring the statistics and then using the standard formulas:
\begin{eqnarray}
T_{SS}({\bf k},{\bf k^{\prime}};E) &=& \frac{1}{2} \left ( {\widehat{T}}_{SS}({\bf k},{\bf k^{\prime}};E) -{\widehat{T}}_{SS}(-{\bf k},{\bf k^{\prime}};E) \right ) \quad \quad (SS \rightarrow VV) \, , \nonumber \\
T_{SV}({\bf k},{\bf k^{\prime}};E) &=& \frac{1}{2} \left ( {\widehat{T}}_{SV}({\bf k},{\bf k^{\prime}};E) +{\widehat{T}}_{SV}(-{\bf k},{\bf k^{\prime}};E) \right ) \quad \quad (SV \rightarrow VS) \, .
\label{(2.7)}
\end{eqnarray}

However, the LS equation for $\widehat{T}_{xy}$ is not well defined in any channel and hence using (\ref{(2.5)}) or (\ref{(2.6)})-(\ref{(2.7)}) may not be totally equivalent. In particular, for the SS and VV channels, the UV divergences one finds using (\ref{(2.5)}) are softer than those from (\ref{(2.6)})-(\ref{(2.7)}), so we shall work with (\ref{(2.5)}). For the SV and VS channels, however, the UV divergences found using (\ref{(2.5)}) are as strong as the ones that stem from (\ref{(2.6)})-(\ref{(2.7)}). For convenience, we have chosen to work with the latter for these channels. 

The LS equation in the isoscalar-scalar channel in (\ref{(2.5)}) is already  well defined, as it is apparent from the anti-symmetrization of the corresponding potential (\ref{(2.2)}). On the contrary, the other three channels require regularization. Searching for the systematics to tackle them will be the aim of the next three sections. For notation simplicity, the energy dependence of the T-matrices as well as of other auxiliary functions will not be displayed explicitely for the rest of the paper.

%%%%%%%%%%%%%%%%%%%%%%%%%% ISOVECTOR-SINGLET %%%%%%%%%%%%%%%%%%%%%%%%%%%%%%%%
\section{The isovector-singlet channel}
\indent

The LS equation for this channel reads:
\begin{center}
${\widehat{T}}_{VS}^{ij} ({\bf k}, {\bf k^{\prime}}) = V_{VS}^{ij} ({\bf k},{\bf k^{\prime}})+ \int^{\Lambda} \, \frac{d^{3}k^{\prime \prime}}{(2\pi)^{3}} \, V_{VS}^{ik}({\bf k}, {\bf k^{\prime \prime}}) \, \frac{1}{E-\frac{{\bf k}^{\prime \prime 2}}{M}+i\eta} \, {\widehat{T}}_{VS}^{kj}({\bf k^{\prime \prime}}, {\bf k^{\prime}}) \, ,$ 
\end{center}
\begin{center}
where 
\begin{eqnarray}
V_{VS}^{ij}({\bf k},{\bf k^{\prime}}) &=& \left \{ c_{0}+ \frac{c_2}{({\bf k}-{\bf k^{\prime}})^2+m_{\pi}^2} \right \} \delta^{ij} \, , \nonumber   \\
c_0 &:= & C_S-3C_T+ \left ( \frac{g_A}{2f_{\pi}} \right )^2 \, , \nonumber \\
c_2 &:= & -\left ( \frac{g_Am_{\pi}}{2f_{\pi}} \right )^2 \, ,
\label{(3.1)}
\end{eqnarray}
\end{center}
where in the last lines we remember the values those constants would take if the potential had been calculated at LO in $\chi$PT. The hat and the VS subscript will be dropped in the following.

Let us define:
\begin{eqnarray} 
{\mathcal{A}}({\bf k^{\prime}})\, \delta^{ij} &:= &  \int^{\Lambda } \, \frac{d^3 k^{\prime \prime }}{(2\pi )^3} \, \frac{ T^{ij} ({\bf k^{\prime \prime }},{\bf k^{\prime}} )}{E-\frac{{\bf k}^{\prime \prime 2}}{M}+i\eta } \, . 
\end{eqnarray}

Then (\ref{(3.1)}) reads:   
\begin{eqnarray}
T^{ij} ({\bf k}, {\bf k^{\prime}}) = c_0 (1+{\mathcal{A}}({\bf k^{\prime}}))\, \delta^{ij}+\frac{c_2 \, \delta^{ij}}{({\bf k}-{\bf k^{\prime}})^2+m_{\pi}^2}+\int^{\Lambda} \frac{d^3k^{\prime \prime}}{(2\pi )^3} \, \frac{c_2}{({\bf k}-{\bf k^{\prime \prime}})^2+m_{\pi}^2}\frac{T^{ij}({\bf k}^{\prime \prime}, {\bf k^{\prime}})}{E-\frac{{\bf k}^{\prime \prime 2}}{M}+i\eta } \;  
\label{(3.2)}
\end{eqnarray}
and can be rewritten after solving:
\begin{eqnarray}
T_2 ({\bf k},{\bf k^{\prime}}) &=& \frac{1}{({\bf k}-{\bf k^{\prime}})^2+m_{\pi}^2}+\int^{\Lambda} \frac{d^3k^{\prime \prime}}{(2\pi )^3} \, \frac{c_2}{({\bf k}-{\bf k^{\prime \prime}})^2+m_{\pi}^2}\frac{T_2 ({\bf k}^{\prime \prime}, {\bf k^{\prime}})}{E-\frac{{\bf k}^{\prime \prime 2}}{M}+i\eta } \, ,
\label{(3.2.1)}
\end{eqnarray}
in the form:
\begin{eqnarray}
T ({\bf k}, {\bf k^{\prime}}) &=& c_2 \, T_2({\bf k},{\bf k^{\prime}})+ c_0(1+{\mathcal{A}}({\bf k^{\prime}}))\, \left [ 1 + c_2 \, \int^{\Lambda}\frac{d^3k^{\prime \prime}}{(2\pi )^3} \, \frac{T_2({\bf k},{\bf k^{\prime \prime}})}{E-\frac{{\bf k}^{\prime \prime 2}}{M}+i\eta } \right ] \, ,
\label{(3.2.2)}
\end{eqnarray}
where we have dropped the $\delta^{ij}$ structure.
If ${\mathcal{A}}({\bf k^{\prime}})$ was a fixed function, the equation above would be well defined and could already be solved with no need to regularize it.
 However ${\mathcal{A}}({\bf k^{\prime}})$ is a  functional of $T$ and a second equation which relates them must be introduced. This is achieved by multiplying
eq.(\ref{(3.2.2)}) by $1/(E-\frac{{\bf k}^2}{M}+i\eta)$ and integrating over ${\bf k}$. We obtain:
\begin{eqnarray}
c_0(1+{\mathcal{A}}({\bf k^{\prime}})) &=& \frac{ 1+ c_2 \int^{\Lambda}\frac{d^3k}{(2\pi )^3} \, \frac{T_2({\bf k},{\bf k^{\prime}})}{E-\frac{{\bf k}^2}{M}+i\eta }}{\frac{1}{c_0}- \left [\, {\mathcal{I}}_0+c_2 \, \int^{\Lambda} \frac{d^3k}{(2\pi )^3} \int^{\Lambda} \frac{d^3k^{\prime \prime}}{(2\pi )^3} \frac{1}{E-\frac{{\bf k}^2}{M}+i\eta }T_2({\bf k},{\bf k^{\prime \prime}}) \frac{1}{E-\frac{{\bf k}^{\prime \prime 2}}{M}+i\eta}\, \right ] } \, , \nonumber \\
&& {\mathcal{I}}_0 := \int^{\Lambda} \frac{d^3 k}{(2\pi )^3} \, \frac{1}{E-\frac{{\bf k}^2}{M}+i\eta } \, .
\label{(3.3)}
\end{eqnarray}

Substituting iteratively $T_2$ in (\ref{(3.2.1)}) in the rhs of (\ref{(3.3)}) we see that only the first iteration produces 
further divergent expressions when $\Lambda \rightarrow \infty$. We can then write (\ref{(3.3)}) as:
\begin{eqnarray}
\label{(3.6)}
c_0(1+{\mathcal{A}}({\bf k^{\prime}})) &=& \frac{1+c_2{\mathcal{F}}({\bf k^{\prime}})}{\frac{1}{c_0}- \left [ \, {\mathcal{I}}_0+c_2 \, {\mathcal{L}}+c_2{\mathcal{F}}^{ \, \prime}  \, \right ] } \, ,
\end{eqnarray}
where ${\mathcal{I}}_0$ and ${\mathcal{L}}$ contain linearly and logarithmically divergent terms respectively, whereas ${\mathcal{F}}$ (${\mathcal{F}}^{ \, \prime \, }$) just denote finite functions:
\begin{eqnarray}
{\mathcal{L}} &:=& \int^{\Lambda} \frac{d^3 k}{(2\pi )^3} \int^{\Lambda} \frac{d^3 k^{\prime \prime}}{(2\pi)^3} \frac{1}{E-\frac{{\bf k}^2}{M}+i\eta} \frac{1}{({\bf k}-{\bf k^{\prime \prime}})^2+m_{\pi}^2} \frac{1}{E-\frac{{\bf k^{\prime \prime }}^2}{M}+i\eta} \, , \nonumber \\
{\mathcal{F}}({\bf k^{\prime}}) &:=& \int^{\Lambda} \frac{d^3 k}{(2\pi )^3} \frac{T_2({\bf k},{\bf k^{\prime}})}{E-\frac{{\bf k}^2}{M}+i\eta} \, , \nonumber \\
{\mathcal{F}}^{\, \prime} &:=& \int^{\Lambda} \frac{d^3 k}{(2\pi )^3} \int^{\Lambda} \frac{d^3 k^{\prime \prime}}{(2\pi)^3} \int^{\Lambda} \frac{d^3 k^{\prime \prime \prime }}{(2\pi)^3} \frac{1}{E-\frac{{\bf k}^2}{M}+i\eta} \frac{c_2}{({\bf k}-{\bf k^{\prime \prime}})^2+m_{\pi}^2}\frac{T_2 ({\bf k}^{\prime \prime}, {\bf k^{\prime \prime \prime }})}{E-\frac{{\bf k}^{\prime \prime 2}}{M}+i\eta } \frac{1}{E-\frac{{\bf k^{\prime \prime \prime }}^2}{M}+i\eta} \,  \quad \quad
\end{eqnarray}

It is clear that the expression (\ref{(3.6)}) can be renormalized by a redefinition of $c_0$. In dimensional regularization, (D=3+2$\epsilon $), we obtain:
\begin{eqnarray}
\frac{1}{c_0} &=& -\frac{M^2\, c_2}{4(4\pi)^2}\left ( \frac{1}{\epsilon } + \chi_{sch} \right )+ \frac{1}{c_0^r(\mu )} \, , \nonumber \\ 
\chi_{MS} &=& 0 \, ,\nonumber \\
\chi_{\overline{MS}} &=& \gamma_E - {\rm{Log}}(4\pi ) \, ,
\end{eqnarray}
which is in agreement with \cite{Kaplan}, and 
for a hard cut-off:
\begin{eqnarray}
\frac{1}{c_0} &=& -\frac{M\Lambda }{2\pi^2}+\frac{M^2 \, c_2}{32\pi^2}\, {\rm{Log}}\,\left ( \frac{ \Lambda^2}{\mu^2} \right )+ \frac{1}{c_0^r(\mu )} \, .
\label{(3.7)}
\end{eqnarray}

If we now wish to solve numerically the LS equation, we should proceed as 
usual and introduce a hard cut-off. However $c_0$ is not to be fitted to the experimental data but substituted by (\ref{(3.7)}) and the cut-off made as large as possible (in practice 
it should be enough if $\sqrt{EM}/\Lambda$ is of the order of neglected subleading contributions from the NLO potential, (see 
\cite{Lepage} for a more technical discussion). What we 
have just proved is that the result will be cut-off independent up to corrections  $\sqrt{EM}/\Lambda$. $\mu$ must be fixed at the 
relevant momentum scale $\mu\sim ( \sqrt{EM}, m_{\pi} )$ and $c_0^r(\mu)$ tuned to fit the experimental data. 

Although we have no prediction for $c_0^r(\mu)$ we can try to understand from (\ref{(3.7)}) how large scattering lengths may arise. Since $c_0^r(\mu)$ evolves according to a non-perturbative renormalization group (RG) equation it might take very different values depending on the scale it is evaluated at. After solving it:
\begin{eqnarray}
c_0^r(\mu)&=& \frac{c_0^r(\mu_0)}{1+\frac{M^2c_2c_0^r(\mu_0)}{16\pi^2}{\rm{Log}}\frac{\mu}{\mu_0} } \, .
\end{eqnarray}
if we input the value of ref. \cite{Kaplan} $c_0^r(m_{\pi})=-(\frac{1}{79 \; {\rm{MeV}}})^2$, we obtain $c_0^r(M)=-(\frac{1}{125 \;{\rm{MeV}}})^2$, which is not quite at the natural scale ($\sim M $). Hence, the non-perturbative low energy dynamics does not seem to be enough to fill the gap between the natural scales and the large scattering lengths. In spite of that, the variation of $c_0^r (\mu )$ from $m_{\pi}$ to $M$ is large enough as to justify a non-perturbative treatment of the OPE in this channel.

%%%%%%%%%%%%%%%%%%%%%%% ISOSINGLET-TRIPLET %%%%%%%%%%%%%%%%%%%%%%%%%%%%%  
\section{The isosinglet-vector channel}
\indent 

The LS equation for this channel reads: 
\begin{center}
${\widehat{T}}_{SV}^{ij} ({\bf k}, {\bf k^{\prime}}) = V_{SV}^{ij} ({\bf k},{\bf k^{\prime}})+ \int^{\Lambda} \, \frac{d^{3}k^{\prime \prime}}{(2\pi)^{3}} \, V_{SV}^{ik}({\bf k}, {\bf k^{\prime \prime}}) \, \frac{1}{E-\frac{{\bf k}^{\prime \prime 2}}{M}+i\eta} \, {\widehat{T}}_{SV}^{kj}({\bf k^{\prime \prime}}, {\bf k^{\prime}}) \, , $ \\
\end{center}
\begin{center}
where  
\begin{eqnarray}
V_{SV}^{ij}({\bf k},{\bf k^{\prime}}) &=& \left \{ c_{0}+ \frac{c_2}{({\bf k}-{\bf k^{\prime}})^2+m_{\pi}^2} \right \} \delta^{ij} + c_1\frac{({\bf k}-{\bf k^{\prime}})^i({\bf k}-{\bf k^{\prime}})^j}{({\bf k}-{\bf k^{\prime}})^2+m_{\pi}^2} \, , \nonumber \\
c_0 &:= & C_S+C_T+3 \left ( \frac{g_A}{2f_{\pi}} \right )^2 \, , \nonumber \\
c_1 &:= & -6 \left ( \frac{g_A}{2f_{\pi}} \right )^2 \, , \nonumber \\
c_2 &:= & -3 \left ( \frac{g_Am_{\pi}}{2f_{\pi}} \right )^2 \, ,
\label{(4.1)}
\end{eqnarray}
\end{center}
where we show also the LO values of the coupling constants. 
We shall drop the subscript $SV$ and the hat in the following. We call the term
proportional to $c_1$ above
spin symmetry breaking (SSB) term. This term breaks orbital angular momentum conservation and makes the analysis of
this channel qualitatively different from the previous one. 
In order to illustrate it, let us take ${\bf k^{\prime}}={\bf 0}$ for simplicity. As we regulate (\ref{(4.1)}), the possible divergences arising when the regulator is removed depend on the high momentum behavior of $T^{ij}({\bf k})$. If $T^{ij}({\bf k})
\sim \vert {\bf k}\vert^{\a}$ , the usual power counting arguments imply that, due to the SSB term, 
the integral on the rhs will rise this power by one. Hence, the high momentum 
behavior of the lhs of the equation will not match the one of its rhs unless: (i) $\a =-1$ and the high momentum contribution of the potential
cancels out the one arising from the integral or (ii) $\a=0$ and the bare coupling constant $c_1$ goes to zero as the cut-off goes to infinity, which removes the $\vert {\bf k}\vert^{\a +1}$ term on the rhs.
  We prove in the Appendix A that 
the case (i) in fact reduces to (ii). 

The preceding discussion provides a rather intuitive introduction to what, in the course of section 4.1, we will demonstrate in full detail. That is, all those rising divergences caused by the SSB term can only be renormalized by a, so far undetermined, flowing of their accompanying coupling constant, $c_1$. Next we will fix this cut-off dependence and, having explored the consequences such a behavior has on the amplitude, will come back in section 4.2 to standard procedures. There it is shown that the alternative of treating SSB as a perturbation solves the problem, as all divergences get renormalized by local counterterms and no $c_1$ flowing is longer required.  

%So in section 4.1 we will proceed keeping in mind that $c_1\rightarrow 0$ in s%ome, at the moment undetermined, way. After having explored the consequences t%his flow has on the amplitude, we will examine in section 4.2 the alternative %of treating SSB as a perturbation. In such a case we find that $c_1$ is not re%quired to depend on the cut-off.

\renewcommand{\theequation}{\thesubsection.\arabic{equation}}
\subsection{Non-perturbative treatment of the SSB term}
\setcounter{equation}{0}
\indent 

Let us then return to equation (\ref{(4.1)}). It has the following structure:
\begin{eqnarray}
\label{(4.2)}
T^{ij}({\bf k},{\bf k^{\prime}}) &=& c_0 (\delta^{ij}+{\mathcal{A}}^{ij}({\bf k^{\prime}})) +c_1 \left [  \frac{({\bf k}-{\bf k^{\prime}})^i ({\bf k}-{\bf k^{\prime}})^j}{({\bf k}-{\bf k^{\prime}})^2+m_{\pi}^2}+ {\mathcal{B}}^{ij}({\bf k},{\bf k^{\prime}}) \right ]+ \nonumber \\
&& + c_2\frac{\delta^{ij}}{({\bf k}-{\bf k^{\prime}})^2+m_{\pi}^2} + c_2\int^{\Lambda}\frac{d^3 k^{\prime \prime}}{(2\pi )^3} \frac{1}{({\bf k}-{\bf k^{\prime \prime}})^2+m_{\pi}^2}\frac{T^{ij}({\bf k^{\prime \prime}},{\bf k^{\prime}})}{E-\frac{{\bf k^{\prime \prime}}^2}{M}+i\eta} \, , \nonumber \\
{\mathcal {A}}^{ij}({\bf k^{\prime}}) &=& \int^{\Lambda}\frac{d^3 k}{(2\pi )^3}\frac{T^{ij}({\bf k},{\bf k^{\prime}})}{E-\frac{{\bf k}^2}{M}+i\eta} \, , \nonumber \\
{\mathcal B}^{ij}({\bf k},{\bf k^{\prime}}) &=& \int^{\Lambda}\frac{d^3 k^{\prime \prime}}{(2\pi )^3}\frac{({\bf k}-{\bf k^{\prime \prime}})^i({\bf k}-{\bf k^{\prime \prime}})^k}{({\bf k}-{\bf k^{\prime \prime}})^2+m_{\pi}^2} \frac{T^{kj}({\bf k}^{\prime \prime},{\bf k^{\prime}})}{E-\frac{{\bf k^{\prime \prime}}^2}{M}+i\eta} \, . 
%\nonumber
\end{eqnarray}

Let us define:
\begin{eqnarray}
\label{(4.4)}
T^{ij} ({\bf k},{\bf k^{\prime}}) &:= & c_0(\delta^{ij}+{\mathcal{A}}^{ij}({\bf k^{\prime}})) \, T_0 ({\bf k}) +c_1 \, T_1^{ij}({\bf k},{\bf k^{\prime}}) +c_2 \, T_2 ({\bf k},{\bf k^{\prime}}) \, \delta^{ij} \, , \nonumber \\
T_0 ({\bf k}) &=&  1 + c_2 \int^{\Lambda}\frac{d^3 k^{\prime \prime}}{(2\pi)^3}\frac{1}{({\bf k}-{\bf k^{\prime \prime}})^2+m_{\pi}^2}\frac{T_0 ({\bf k^{\prime \prime}})}{E-\frac{{\bf k^{\prime \prime}}^2}{M}+i\eta} \, , \nonumber \\
T_1^{ij}({\bf k},{\bf k^{\prime}}) &=& \frac{({\bf k}-{\bf k^{\prime}})^i ({\bf k}-{\bf k^{\prime}})^j}{({\bf k}-{\bf k^{\prime}})^2+m_{\pi }^2}+{\mathcal{B}}^{ij}({\bf k},{\bf k^{\prime}})+c_2 \int^{\Lambda}\frac{d^3 k^{\prime \prime}}{(2\pi)^3}\frac{1}{({\bf k}-{\bf k^{\prime \prime}})^2+m_{\pi}^2}\frac{T_1^{ij}({\bf k^{\prime \prime}},{\bf k^{\prime}})}{E-\frac{{\bf k^{\prime \prime}}^2}{M}+i\eta} \, , \nonumber \\
T_2 ({\bf k},{\bf k^{\prime}}) &=& \frac{1}{({\bf k}-{\bf k^{\prime}})^2+m_{\pi }^2}+ c_2 \int^{\Lambda}\frac{d^3 k^{\prime \prime}}{(2\pi)^3}\frac{1}{({\bf k}-{\bf k^{\prime \prime}})^2+m_{\pi}^2}\frac{T_2 ({\bf k^{\prime \prime}},{\bf k^{\prime}})}{E-\frac{{\bf k^{\prime \prime}}^2}{M}+i\eta} \, ,
\end{eqnarray}
which allows us to isolate in $T_1^{ij}({\bf k},{\bf k^{\prime}})$ and $c_0 \, (\delta^{ij}+ {\mathcal{A}}^{ij}({\bf k^{\prime}}) )$ all sources of divergent behavior, since $T_0({\bf k})$ and $T_2({\bf k},{\bf k^{\prime}})$ are perfectly well defined.

Using the expressions of ${\mathcal{B}}^{ij}({\bf k},{\bf k^{\prime}})$ in (\ref{(4.2)}) and $T^{ij}({\bf k},{\bf k^{\prime}})$ in (\ref{(4.4)}), $T_1^{ij}({\bf k},{\bf k^{\prime}})$ can be re-casted in the form:
\begin{eqnarray}
T_1^{ij}({\bf k},{\bf k^{\prime}}) &=& c_0 (\delta^{kj}+{\mathcal{A}}^{kj}({\bf k^{\prime}}))\, T_{10}^{ik}({\bf k}) + T_{11}^{ij}({\bf k},{\bf k^{\prime}})+ c_2 T_{12}^{ij}({\bf k},{\bf k^{\prime}})\, , \nonumber \\
T_{10}^{ij}({\bf k}) &=& \int^{\Lambda} \frac{d^3 k^{\prime \prime}}{(2\pi)^3} \frac{({\bf k}-{\bf k^{\prime \prime}})^i({\bf k}-{\bf k^{\prime \prime}})^j}{({\bf k}-{\bf k^{\prime \prime}})^2+m_{\pi}^2} \frac{T_0 ({\bf k}^{\prime \prime})}{E-\frac{{\bf k^{\prime \prime}}^2}{M}+i\eta}+ \nonumber \\
&& + \int^{\Lambda} \frac{d^3 k^{\prime \prime}}{(2\pi)^3} \frac{c_1({\bf k}-{\bf k^{\prime \prime}})^i({\bf k}-{\bf k^{\prime \prime}})^k +c_2 \, \delta^{ik}}{({\bf k}-{\bf k^{\prime \prime}})^2+m_{\pi}^2} \frac{T_{10}^{kj}({\bf k}^{\prime \prime})}{E-\frac{{\bf k^{\prime \prime}}^2}{M}+i\eta} \, , \nonumber \\
T_{11}^{ij}({\bf k},{\bf k^{\prime}}) &=& \frac{({\bf k}-{\bf k^{\prime}})^i ({\bf k}-{\bf k^{\prime}})^j}{({\bf k}-{\bf k^{\prime}})^2+m_{\pi }^2}+ \int^{\Lambda}\frac{d^3 k^{\prime \prime}}{(2\pi)^3} \frac{c_1({\bf k}-{\bf k^{\prime \prime}})^i({\bf k}-{\bf k^{\prime \prime}})^k +c_2 \, \delta^{ik}}{({\bf k}-{\bf k^{\prime \prime}})^2+m_{\pi}^2} \frac{T_{11}^{kj}({\bf k}^{\prime \prime},{\bf k^{\prime}})}{E-\frac{{\bf k^{\prime \prime}}^2}{M}+i\eta} \, , \nonumber \\
T_{12}^{ij}({\bf k},{\bf k^{\prime}}) &=&\int^{\Lambda} \frac{d^3 k^{\prime \prime}}{(2\pi)^3} \frac{({\bf k}-{\bf k^{\prime \prime}})^i({\bf k}-{\bf k^{\prime \prime}})^j}{({\bf k}-{\bf k^{\prime \prime}})^2+m_{\pi}^2} \frac{T_2 ({\bf k}^{\prime \prime},{\bf k^{\prime}})}{E-\frac{{\bf k^{\prime \prime}}^2}{M}+i\eta}+ \nonumber \\
&& + \int^{\Lambda} \frac{d^3 k^{\prime \prime}}{(2\pi)^3} \frac{c_1({\bf k}-{\bf k^{\prime \prime}})^i({\bf k}-{\bf k^{\prime \prime}})^k +c_2 \, \delta^{ik}}{({\bf k}-{\bf k^{\prime \prime}})^2+m_{\pi}^2} \frac{T_{12}^{kj}({\bf k}^{\prime \prime},{\bf k^{\prime}})}{E-\frac{{\bf k^{\prime \prime}}^2}{M}+i\eta} \, .
\label{(4.7)}
\end{eqnarray}

This decomposition enables us to compute  $c_0(\delta^{ij}+{\mathcal{A}}^{ij}({\bf k^{\prime}}))$, and hence the full amplitude $T^{ij}({\bf k},{\bf k^{\prime}})$, in terms of $T_0({\bf k})$, $T_{1n}^{ij}({\bf k},{\bf k^{\prime}})$ ($n=0,1,2$) and $T_{2}({\bf k},{\bf k^{\prime}})$ through the equation 

\begin{eqnarray}
\left [ \, \frac{\delta^{ik}}{c_0}-\int^{\Lambda}\frac{d^3 k}{(2\pi)^3} \frac{T_0 ({\bf k})\, \delta^{ik}}{E-\frac{{\bf k}^2}{M}+i\eta}-c_1 \int^{\Lambda}\frac{d^3 k}{(2\pi)^3} \frac{T_{10}^{ik}({\bf k})}{E-\frac{{\bf k}^2}{M}+i\eta} \, \right ] \, c_0 (\delta^{kj}+{\mathcal{A}}^{kj}({\bf k^{\prime}})) = \nonumber \\
= \delta^{ij}+c_1\int^{\Lambda} \frac{d^3 k}{(2\pi)^3} \frac{T_{11}^{ij}({\bf k},{\bf k^{\prime}})+c_2T_{12}^{ij}({\bf k},{\bf k^{\prime}})}{E-\frac{{\bf k}^2}{M}+i\eta}+c_2\int^{\Lambda}\frac{d^3 k}{(2\pi)^3} \frac{T_2 ({\bf k},{\bf k^{\prime}}) \, \delta^{ij}}{E-\frac{{\bf k}^2}{M}+i\eta} \, .  
\end{eqnarray}

As we have already mentioned  $T_0({\bf k})$ and $T_2 ({\bf k},{\bf k^{\prime}})$ are finite when the cut-off is removed.
If we solve $T_{1n}^{ij}({\bf k},{\bf k^{\prime}})$, $n=0, 1, 2$ 
iteratively,
the most divergent pieces in the $n$-th iteration are $T_{10}\sim (c_1 \Lambda)^{n}\Lambda $, $T_{11}\sim (c_1 \Lambda)^{n}$ and $T_{12}\sim (c_1 \Lambda)^{n-1}c_1$. These series are expected to have a finite radius of convergence. The radius of convergence is in any case non-zero because they are bounded by geometric series
(or derivatives of them). If $c_1$ does not go to zero as $1/\Lambda$ or stronger (in particular, if $c_1$ is not allowed to flow), each series will separately diverge. In that case, a finite result can only be obtained if non-trivial cancellations occur for all $n$, which we do not see how they could actually
 happen.
If, on the contrary,
\begin{eqnarray}
c_1(\Lambda) &=& {{\bar c}_1 \over \Lambda} + ... \; . 
\label{(4.9)}
\end{eqnarray}
and ${\bar c}_1$ is small enough, the series will converge. For the T-matrix, such a strong cut-off dependence implies that the terms:
\begin{eqnarray}
c_1T_{10}^{ij}({\bf k}) &\longrightarrow & t_{10}^{(0)} \delta^{ij}+\frac{t_{10}^{ij}({\bf k})}{\Lambda}+ \, ...  \, , \nonumber \\ 
c_1T_{11}^{ij}({\bf k},{\bf k^{\prime}}) &\longrightarrow & \frac{t_{11}^{ij}({\bf k},{\bf k^{\prime}})}{\Lambda}+ \, ...  \, , \nonumber \\
c_1T_{12}^{ij}({\bf k},{\bf k^{\prime}}) &\longrightarrow & \frac{t_{12}^{ij}({\bf k},{\bf k^{\prime}})}{\Lambda} + \, ...  \, ,
\end{eqnarray}
where $t_{10}^{(0)}$ is simply a finite constant and, as we see, all ${\bf k}$, ${\bf k^{\prime}}$-encoded information will be washed out from the amplitude.

That is to say:
\begin{eqnarray}
T^{ij}({\bf k},{\bf k^{\prime}}) &=& \lim_{\Lambda \rightarrow \infty}  c_0 (\delta^{kj}+ {\mathcal{A}}^{kj}({\bf k^{\prime}}))\left ( T_0 ({\bf k}) \delta^{ik}+c_1T_{10}^{ik}({\bf k})\right ) +c_1T_{11}^{ij}({\bf k},{\bf k^{\prime}})+c_2 \left ( T_2 ({\bf k},{\bf k^{\prime}}) \, \delta^{ij} + \right. \nonumber \\
&& \left. + c_1T_{12}^{ij}({\bf k}, {\bf k^{\prime}}) \right )  = c_0 (\delta^{ij}+{\mathcal{A}}^{ij}({\bf k^{\prime}})) \left (T_0 ({\bf k})+c_1\, t_{10}^{(0)} \right )+c_2 T_2 ({\bf k},{\bf k^{\prime}})\delta^{ij} \, ,
\label{(4.11)}
\end{eqnarray}
which is finite provided $c_0 \, (\delta^{ij}+{\mathcal{A}}^{ij}({\bf k^{\prime}}))$ is finite. In order to prove the latter we borrow from section 3 the following results:
\begin{eqnarray} 
\int^{\Lambda} \frac{d^3k}{(2\pi)^3}\frac{T_0 ({\bf k})}{E-{k^2\over M}+i\eta} &=& -\frac{M\Lambda}{2\pi^2}+\frac{M^2c_2}{32\pi^2}{\rm{Log}}\left ( \frac{\Lambda^2}{\mu^2} \right )+ {\mathcal{O}}\left ( 1 \right ) \, , \nonumber \\
\int^{\Lambda} \frac{d^3k}{(2\pi)^3} \frac{T_2 ({\bf k},{\bf k^{\prime}})}{E-{k^2\over M}+i\eta} &=& {\mathcal{O}} \left ( 1 \right ) \, ,
\end{eqnarray}
and find in Appendix B:
\begin{eqnarray}
\label{(4.12)}
c_1 \int^{\Lambda} \frac{d^3k}{(2\pi)^3} \frac{ T_{10}^{ii}({\bf k})} {E-{k^2\over M}+i\eta} &=& a_0\Lambda + ib_0 \sqrt{EM}+ 
d_0{\rm{Log}}\left ( {\Lambda\over m_{\pi}} \right )+{\mathcal{O}}\left ( {1\over \Lambda}\right ) \, , \nonumber \\
c_1 \int^{\Lambda} \frac{d^3k}{(2\pi)^3} \frac{T_{11}^{ii}({\bf k},{\bf k^{\prime}})} {E-{k^2\over M}+i\eta} &=& {\mathcal{O}}\left ({1} \right ) \, , \nonumber \\
c_1 \int^{\Lambda} \frac{d^3k}{(2\pi)^3} \frac{T_{12}^{ii}({\bf k},{\bf k^{\prime}})} {E-{k^2\over M}+i\eta} &=& {\mathcal{O}} \left ( {1\over \Lambda} \right ) \, ,
\end{eqnarray}
where $a_0$, $b_0$, $d_0$ are cut-off independent constants related to ${\bar c}_1$. Then the flow:
\begin{eqnarray}
{1\over c_0} &=& -\frac{M\Lambda}{2\pi^2}+\frac{a_0\Lambda}{3}+\frac{M^2c_2}{32\pi^2} {\rm{Log}}\left ( \frac{ \Lambda^2}{\mu^2} \right )+\frac{d_0}{6}{\rm{Log}}\left ( \frac{\Lambda^2}{\mu^2} \right )+ \frac{1}{c_0^r(\mu )} \, ,
\label{(4.13)}
\end{eqnarray}
makes $c_0(\delta^{ij}+{\mathcal{A}}^{ij}({\bf k^{\prime}}))$ finite and hence does (\ref{(4.11)}). We have then proved that the 
flows (\ref{(4.9)}) and (\ref{(4.13)}) renormalize the triplet channel. 

It is not difficult to see that the various series above involving divergent terms are bounded by geometric series or derivatives of them. This ensures that our flows provide actually finite expressions for the amplitude if $\bar{c}_1$ is small enough. However, this amplitude appears to be diagonal in spin space
and hence orbital angular momentum is conserved. Although, the observed 
$^3S_1$-$^3D_1$  mixing, which is small, might be attributed to a higher order 
effect, it is clear from ref. \cite{Kaiserhw} that it
is due to the OPE to a large extend.
In order to preclude the conservation of orbital angular momentum, we can foresee two ways out: (i) a 
SSB term may survive in the renormalized amplitude if $\bar{c}_1$ is tuned 
infinitely close to the radius of convergence of the series, so that our bounds do not hold anymore, and (ii) the SSB term from OPE must be treated as a perturbation and renormalized as such.
The possibility (i) is examined in Appendix C, where we show it unlikely to be realized. In the following subsection we explore (ii) and 
prove that if a suitable SSB term is treated as a perturbation, the amplitude is renormalizable at first order and the mixing survives.
       
\subsection{Treating the SSB term perturbatively}
\setcounter{equation}{0}
\indent

Let us split the potential as:
\begin{eqnarray}
V^{ij} ({\bf k},{\bf k^{\prime }}) &=& V^{(0) \, ij}({\bf k},{\bf k^{\prime }})+V^{(1) \, ij}({\bf k},{\bf k^{\prime }}) \, , \nonumber \\
V^{(0) \, ij}({\bf k},{\bf k^{\prime }}) &=& \left \{  {\widetilde{c}}_0 + \frac{{\widetilde{c}}_2}{({\bf k}-{\bf k^{\prime}})^2+m_{\pi}^2} \right \} \delta^{ij} \, , \nonumber \\
V^{(1)\, ij}({\bf k},{\bf k^{\prime }}) &=& {\widetilde{c}}_1 \frac{({\bf k}-{\bf k^{\prime}})^i({\bf k}-{\bf k^{\prime}})^j-\frac{({\bf k}-{\bf k^{\prime}})^2}{3}\delta^{ij}}{({\bf k}-{\bf k^{\prime}})^2+m_{\pi}^2}  \, , \nonumber \\
{\widetilde{c}}_0 &:=& C_S + C_T + \left (\frac{g_A}{2f_{\pi}} \right )^2 \, , \nonumber \\
{\widetilde{c}}_1 &:=& -6 \left ( \frac{g_A}{2f_{\pi}} \right )^2 \, , \nonumber \\
{\widetilde{c}}_2 &:=& -\left ( \frac{g_A m_{\pi}}{2f_{\pi}} \right )^2 \, ,  
\end{eqnarray}
with LO values for the coupling constants indicated. In the following we drop the SV-channel sub-indexes. 

The amplitude will be written as:
\begin{eqnarray}
T^{ij}({\bf k},{\bf k^{\prime}}) &=& T^{(0) \, ij}({\bf k},{\bf k^{\prime }})+T^{(1) \, ij}({\bf k},{\bf k^{\prime }}) \, ,
\end{eqnarray}
where $T^{(0)\, ij}({\bf k},{\bf k^{\prime}})$ fulfills:
\begin{eqnarray}
T^{(0)\, ij}({\bf k},{\bf k^{\prime }}) = V^{(0)\, ij}({\bf k},{\bf k^{\prime }}) + \int^{\Lambda} \frac{d^3 k^{\prime \prime}}{(2\pi)^3} V^{(0)\, ik}({\bf k},{\bf k^{\prime \prime }}) \frac{1}{E-\frac{{\bf k^{\prime \prime}}^2}{M}+i\eta} T^{(0)\, kj}({\bf k^{\prime \prime}},{\bf k^{\prime}}) \, . \,
\end{eqnarray}

The renormalized solution to this equation is given 
by $T^{(0)\, ij}({\bf k},{\bf k^{\prime }})=T({\bf k},{\bf k^{\prime }})\delta^{ij}$
in section 3. At first order in perturbation theory $T^{(1)\, ij}({\bf k},{\bf k^{\prime }})$ verifies:
\begin{eqnarray}
T^{(1) \, ij}({\bf k},{\bf k^{\prime }}) &=& V^{(1)\, ij}({\bf k},{\bf k^{\prime }})+ \int^{\Lambda} \frac{d^3 k^{\prime \prime}}{(2\pi)^3} V^{(1)\, ik}({\bf k},{\bf k^{\prime \prime }}) \frac{1}{E-\frac{{\bf k^{\prime \prime}}^2}{M}+i\eta} T^{(0)\, kj}({\bf k^{\prime \prime}},{\bf k^{\prime}})+ \nonumber \\
&& +\int^{\Lambda} \frac{d^3 k^{\prime \prime}}{(2\pi)^3} V^{(0)\, ik}({\bf k},{\bf k^{\prime \prime }}) \frac{1}{E-\frac{{\bf k^{\prime \prime}}^2}{M}+i\eta} T^{(1)\, kj}({\bf k^{\prime \prime}},{\bf k^{\prime}}) \, .
\end{eqnarray}

Using (\ref{(3.2.1)}) and (\ref{(3.2.2)}) we can see that the second term above is finite. We can then gather the first and second terms into a new, energy dependent, potential defined as:
\begin{eqnarray}
\widetilde{V}^{(1)\, ij}({\bf k},{\bf k^{\prime \prime}}) := V^{(1)\, ij}({\bf k},{\bf k^{\prime }})+ \int^{\Lambda} \frac{d^3 k^{\prime \prime}}{(2\pi)^3} V^{(1)\, ik}({\bf k},{\bf k^{\prime \prime }}) \frac{1}{E-\frac{{\bf k^{\prime \prime}}^2}{M}+i\eta} T^{(0)\, kj}({\bf k^{\prime \prime}},{\bf k^{\prime}}) \, . \,
\end{eqnarray}

Therefore, the integral equation reduces to:
\begin{eqnarray}
T^{(1)\, ij}({\bf k},{\bf k^{\prime }}) &=& \widetilde{V}^{(1)\, ij}({\bf k},{\bf k^{\prime \prime}})+{\widetilde{c}}_0 \, {\mathcal{R}}^{ij}({\bf k^{\prime}}) + \int^{\Lambda} \frac{d^3 k^{\prime \prime}}{(2\pi)^3} \frac{{\widetilde{c}}_2}{({\bf k}-{\bf k^{\prime \prime}})^2+m_{\pi}^2} \frac{T^{(1)\, ij}({\bf k^{\prime \prime}},{\bf k^{\prime}})}{E-\frac{{\bf k^{\prime \prime}}^2}{M}+i\eta} \, , \nonumber \\
{\mathcal{R}}^{ij}({\bf k^{\prime}}) &:=& \int^{\Lambda}\frac{d^3 k^{\prime \prime}}{(2\pi)^3}\frac{T^{(1)\, ij}({\bf k^{\prime \prime}},{\bf k^{\prime}})}{E-\frac{{\bf k^{\prime \prime}}^2}{M}+i\eta} \, .
\end{eqnarray}

In order to prove it finite we decompose:
\begin{eqnarray}
T^{(1)\, ij}({\bf k},{\bf k^{\prime }}) &=& {\widetilde{c}}_0 \, {\mathcal{R}}^{kj}({\bf k^{\prime}})\, T_0^{ik}({\bf k})+ {\widetilde{T}}^{ij}_1({\bf k},{\bf k^{\prime}}) \, ,
\end{eqnarray}
with $T_0^{ij}({\bf k})$ defined in (\ref{(4.4)}) and ${\widetilde{T}}^{ij}_1({\bf k},{\bf k^{\prime}})$ given by:
\begin{eqnarray}
{\widetilde{T}}^{ij}_1({\bf k},{\bf k^{\prime}}) &:=& \widetilde{V}^{(1)\, ij}({\bf k},{\bf k^{\prime \prime}}) + \int^{\Lambda} \frac{d^3 k^{\prime \prime}}{(2\pi)^3} \frac{{\widetilde{c}}_2}{({\bf k}-{\bf k^{\prime \prime}})^2+m_{\pi}^2} \frac{{\widetilde{T}}_1^{ij}({\bf k^{\prime \prime}},{\bf k^{\prime}})}{E-\frac{{\bf k^{\prime \prime}}^2}{M}+i\eta} \, .
\end{eqnarray}

Both $T_0^{ij}({\bf k})$ and ${\widetilde{T}}^{ij}_1({\bf k},{\bf k^{\prime}})$ are well defined (the tensor structure is crucial for the latter to be so). Divergences can only arise in ${\widetilde{c}}_0 \, {\mathcal{R}}^{ij}({\bf k^{\prime}})$, which reads:
\begin{eqnarray}
{\widetilde{c}}_0 \, {\mathcal{R}}^{ij}({\bf k^{\prime}}) &=& \frac{\int^{\Lambda}\frac{d^3 k}{(2\pi)^3} \frac{{\widetilde{T}}_1^{ij}({\bf k},{\bf k^{\prime}})}{E-\frac{{\bf k}^2}{M}+i\eta}}{{\widetilde{c}}_0^{ \, -1}- \frac{1}{3}\int^{\Lambda}\frac{d^3 k}{(2\pi)^3}\frac{T_0^{ii}({\bf k})}{E-\frac{{\bf k}^2}{M}+i\eta}} \, .
\end{eqnarray}

The numerator is well defined (for that the tensor structure is again crucial) and the divergences in the denominator have exactly the same structure as in the denominator of (\ref{(3.6)}). Hence they are renormalized by the same $c_0$ flows. We have then proved that if we treat the SSB term as a perturbation, the amplitude is renormalizable at first order in perturbation theory and no extra counterterm needs to be introduced.
%%%%%%%%%%%%%%%%%%%%%%%%%% ISOVECTOR-VECTOR %%%%%%%%%%%%%%%%%%%%%%%%%%%%%%%%%%
\renewcommand{\theequation}{\thesection \arabic{equation}}
\section{Isovector-vector channel}
\indent

If we use (\ref{(2.6)})-(\ref{(2.7)}) in order to obtain $T_{VV}({\bf k},{\bf k^{\prime}})$, the calculation of ${\widehat{T}}_{VV}({\bf k},{\bf k^{\prime}})$ would reduce to that of the previous section. However, as mentioned in section 2, the UV behavior is smoother in terms of (\ref{(2.5)}), as it happens in the SS channel, although here we still need to introduce a regularization. The LS equation, dropping de isospin delta, reads: 
\begin{eqnarray}
T_{VV}^{ij}({\bf k},{\bf k^{\prime}} ) &=& V_{VV}^{A \, , ij}({\bf k},{\bf k^{\prime}}) + \int^{\Lambda} \frac{d^3k^{\prime \prime}}{(2\pi)^3} V_{VV}^{A,\, ik} ({\bf k},{\bf k^{\prime \prime}}) \frac{1}{E-\frac{{\bf k^{\prime \prime}}^2}{M}+i\eta} T_{VV}^{kj}({\bf k^{\prime \prime}}, {\bf k^{\prime}}) \, , 
\end{eqnarray}
where:
\begin{eqnarray}
V^{A,\, ij}_{VV}({\bf k},{\bf k^{\prime}} ) &=& \frac{1}{2} \left ( V_{VV}^{ij}({\bf k},{\bf k^{\prime}})- V_{VV}^{ij}(- {\bf k},{\bf k^{\prime}}) \right ) = \frac{c_1}{2} \left ( \frac{({\bf k}-{\bf k^{\prime}})^i({\bf k}-{\bf k^{\prime}})^j}{({\bf k}-{\bf k^{\prime}})^2+m_{\pi}^2}-\frac{({\bf k}+{\bf k^{\prime}})^i({\bf k}+{\bf k^{\prime}})^j}{({\bf k}+{\bf k^{\prime}})^2+m_{\pi}^2} \right )+ \nonumber \\
&& + \frac{c_2}{2} \left ( \frac{\delta^{ij}}{({\bf k}-{\bf k^{\prime}})^2+m_{\pi}^2} -\frac{\delta^{ij}}{({\bf k}+{\bf k^{\prime}})^2+m_{\pi}^2} \right ) \, ,
\label{(5.2)}
\end{eqnarray}
where those constants calculated at first order in $\chi$PT take the values:
\begin{eqnarray}
c_1 &:= & 2 \left ( \frac{g_A}{2f_{\pi}} \right )^2 \, , \nonumber \\
c_2 &:= &  \left ( \frac{g_Am_{\pi}}{2f_{\pi}} \right )^2 \, .
\end{eqnarray}

We have not analyzed the possible existence of non-trivial flows which may renormalize the above equation. The fact that the SSB term must be treated perturbatively in the SV channel, indicates that also here we should proceed according to the same philosophy. The potential (\ref{(5.2)}) in the zeroth order approximation reads:
\begin{eqnarray}
V^{(0),\, ij}_{VV}({\bf k},{\bf k^{\prime}}) &=& \frac{c_2}{2} \left ( \frac{\delta^{ij}}{({\bf k}-{\bf k^{\prime}})^2+m_{\pi}^2} -\frac{\delta^{ij}}{({\bf k}+{\bf k^{\prime}})^2+m_{\pi}^2} \right ) \, ,
\end{eqnarray}
which leads to a well defined LS equation. At first order in perturbation theory we will have:
\begin{eqnarray}
T_{VV}^{ij}({\bf k},{\bf k^{\prime}}) &=& T_{VV}^{(0) \, ij}({\bf k},{\bf k^{\prime}}) +  T_{VV}^{(1) \, ij}({\bf k},{\bf k^{\prime}}) \, , \nonumber \\
T_{VV}^{(1) \, ij}({\bf k},{\bf k^{\prime}}) &=& V^{(1)\, ij }_{VV}({\bf k},{\bf k^{\prime}}) + \int^{\Lambda} \frac{d^3 k^{\prime \prime}}{(2\pi )^3} V^{(1) \, ik}_{VV} ({\bf k},{\bf k^{\prime \prime}}) \frac{T_{VV}^{(0) \, kj}({\bf k^{\prime \prime}},{\bf k^{\prime}})}{E-\frac{{\bf k^{\prime \prime}}^2}{M}+i\eta}+ \nonumber \\
&& + \int^{\Lambda} \frac{d^3 k^{\prime \prime}}{(2\pi )^3} V^{(0) \, ik}_{VV} ({\bf k},{\bf k^{\prime \prime}}) \frac{T_{VV}^{(1) \, kj}({\bf k^{\prime \prime}},{\bf k^{\prime}})}{E-\frac{{\bf k^{\prime \prime}}^2}{M}+i\eta} \, ,
\end{eqnarray}
which is also well defined. We expect the divergences arising at higher orders to be absorbed by local counterterms.

%%%%%%%%%%%%%%%%%%%%%DISCUSSION%%%%%%%%%%%%%%%%%%%%%%%%%%%%%%%%%%%%%%%%%

\section{Discussion}
\indent

We have addressed the renormalization of the LS equation for the LO potentials
(in the $\chi$PT counting)
of the NN system in all channels. In addition, for each channel we have been able to carry out our analysis for all partial waves (including partial wave mixing) at once. The isoscalar scalar channel does not require regularization. For the isovector scalar channel we recover the flows of ref. \cite{Kaplan}. The remaining two channels have deserved a more detailed study. 

The first non-trivial result is that the renormalization of the isoscalar vector channel requires a strong flow of the coupling constant of a non-local potential, the SSB one, or, in other words, if only the coupling constants of the local potentials are allowed to flow, the isoscalar vector channel is not renormalizable. Several comments are in order.

First of all, the flow (\ref{(4.9)}) of the coupling constant of the SSB term
is not such a big surprise. Notice that at high momentum this term 
tends to a (direction dependent) constant, which is the same behavior (except for the direction dependence) as the $\delta$-function term both in the singlet and the triplet channel, the coupling constants of which also show similar flows. The main difference is that the leading behavior for $c_0$ is fixed and the subleading contains the free parameter ($c_0^r(\mu)$). For $c_1$ instead,
the leading behavior contains the free parameter (${\bar c}_1$) and the subleading behavior is not observable. 

What is worse, the flow (\ref{(4.9)}) has undesirable consequences:
the renormalized T-matrix conserves orbital angular momentum, even if the bare interaction does not (see Appendix C)
\footnote{ We have also checked perturbatively in ${\bar c}_1$ and $c_2$ up to order
${\bar c}_1c_2$ that the effective range depends on ${\bar c}_1$ only through the scattering length. Since the latter can be adjusted by tuning $c_0^r(\mu)$, up to this order both the scattering length and the effective range are blind to ${\bar c}_1$. We have not looked at what happens to the rest of the amplitude or to higher orders but we suspect that they are also insensitive to ${\bar c}_1$.}. 
Since it is precisely the OPE the main responsible for mixing (also of higher partial waves \cite{Kaiserhw}), we would like it to keep doing this job for us. Therefore, in order for $c_1$ not to flow, but to be fixed at the HB$\chi$L values and produce partial wave mixing, we are forced to exclude the SSB term from the (low energy) LO potential, and to treat it as a perturbation. This also appears to be reasonable
from the phenomenological point of view since the observed mixings are small \cite{Kaiserhw}.  

We have developed this line in sections 4.2 and 5. We have 
proved that at first order the vector channels remain renormalizable (at zeroth order the problem reduces to the one in the singlet channels, which are renormalizable). 
The picture which emerges is half way between \cite{KSW}, where the pions are treated perturbatively, and \cite{Meissner,Kolck} where the whole potential is treated non-perturbatively. The (low energy) LO potential is the part
of the LO potential in the $\chi$PT counting which conserves orbital angular momentum. We are tempted to propose the following counting. The ${\mathcal O}(Q^{n})$ ($n=0,1,...)$ contribution to the NN potential in the $\chi$PT counting must be divided into two
pieces: the one which conserves orbital angular momentum (SS) and the one
which does not (SSB). The SSB terms keep their $\chi$PT counting but the
SS ones are enhanced and 
must be counted as ${\mathcal O}(Q^{n-1})$. Only the LO potential ${\mathcal O}(Q^{-1})$ must be treated (and renormalized) non-perturbatively. We have seen here that this proposal is theoretically consistent at next to leading order, and, in addition, it does not require any coupling constant of a non-local potential to flow anymore. It remains to be seen
if it is still so beyond that order and, of course, whether it 
is phenomenologically successful.

Let us finally comment on recent work on the subject \cite{Bedaque,Frederico}. The authors in both references try to renormalize the triplet channel by adjusting
the coupling constant of the $\delta$-potential only. Hence, according to our results both works should show a remnant cut-off dependence when the cut-off is large enough. Note also that it is only in the large cut-off limit when a meaningful comparison is possible, since
the regularizations used in the three works are different.
The authors of ref. \cite{Frederico}, who use a subtracted ($\mu$-dependent) LS equation, argue that a reasonable boundary condition is that for large $\mu$ the T-matrix coincides with the potential,
and check numerically whether, once the scattering lengths are fixed, the remaining observables are independent of $\mu$ for large $\mu$. They find that for laboratory energies up to $100\;  MeV.$ the
$^3 S_1$ and $^3 D_1$ phase shifts are remarkably independent of $\mu$ 
for $\mu\geq 0.8\; GeV.$ but the mixing angle shows a strong $\mu$-dependence for $6 \; GeV.\geq \mu \geq 0.8 \; GeV.$ and only for $\mu \geq 6\; GeV.$ the $\mu$-dependence smooths and the results 
may appear to converge.
We interpret this stronger $\mu$-dependence of the mixing angle as an 
indication of the remnant cut-off dependence mentioned above. 
The authors of ref. \cite{Bedaque} obtain the flows by analyzing the short distance behavior of the Schr\"odinger equation (see also \cite{Childress}). For the $^1S_0$ they are in qualitative agreement with ours. For the triplet channel they present analytic flow equations which are argued to coincide with those of the chiral limit. The flow of the $\delta$-function term
is given implicitly by their equation (18). They assume that their $\alpha_{\pi}$, which is proportional to our $c_1$, does not flow\footnote{Whereas the combination $\alpha_{\pi}m_{\pi}^2$ that appears in the singlet channel is equivalent to out $c_2$ and, accordingly, remains fixed.} and find a multi-branch
structure for the flow of their $V_0 R^3$, which is proportional to our $c_0$ ($R\rightarrow 0$, $R$ playing the role of an inverse cut-off). 
It is interesting to note that if they allowed $\alpha_{\pi}$ flow like our $c_1$ in section 4.1, namely 
$\alpha_{\pi}\sim R$, and $V_0 R^3$ like our  $c_0$, namely $V_0\sim 1/R^2$,
their eq. (18) becomes cut-off independent. Hence our flow (4.1.5)
is a solution in the $R\rightarrow 0$ limit to the flow equation
(18) of ref. \cite{Bedaque}.
Recall, however, that, if $\alpha_{\pi}$ is not allowed to flow, the
%multi-branch structure prevents the 
strict limit $R\rightarrow 0$ cannot be taken.
%More precisely, it is not difficult to prove that there is no continuous solution to this equation for $R\rightarrow 0$
 This is proved in Appendix D. 
Hence eq. (18) of  \cite{Bedaque} does not produce an acceptable flow for $V_0$ and, therefore, it cannot be used to properly renormalize
%\footnote{We use here the standard meaning of {\it renormalization} in quantum field theory, namely that the cut-off can be taken arbitrarily large (like, for instance, in refs. \cite{ KSW, Kaplan, Frederico}).}
 the triplet channel.
Furthermore, we would like to remark that leaving the cut-off finite and checking that the finite cut-off effects are higher order in the EFT expansion is a procedure which unavoidably 
will lead to problems in this case. If we wish to improve the accuracy of our EFT calculation, we will have to calculate at higher orders. Even if we insist in keeping $R$ finite, we will have to choose it smaller and smaller for the LO terms not to jeopardize the accuracy of the higher order calculation. Then at some point $R$ will hit the region were no continuous solution exists and we will loose all predictive power (if we give up continuity, an infinite number of inequivalent solutions exists). Note that the fact that finite cut-off effects can be compensated by higher dimensional operators \cite{Lepage2}, which holds in perturbatively renormalizable (and asymptotically free) theories, needs not hold here.

%%%%%%%%%%%%%%%%%%%%%%%%%%%%%%%%%%%%%%%%%%%%%%%%%%%%%%%%%%%%%%%%%%%%%%%%%%%%%%%
%%%%%%%%%%%%%%%%%%%%%%%%%%%% ACKNOWLEDGMENTS %%%%%%%%%%%%%%%%%%%%%%%%%%%%%%%%%%

\section*{Acknowledgments}
\indent
We thank Aneesh Manohar and Iain Stewart for discussions in the early stages of this work. We are grateful to Jos\'e Luis Goity for discussions 
and the critical reading of the manuscript. 
Thanks are also given to S. R. Beane, J. Gegelia and D. Phillips for sending their comments to us.
We are supported by the AEN98-031 (Spain) and the 1998SGR 00026 (Catalonia).
D.E. also acknowledges financial support from a MEC FPI fellowship (Spain).

%%%%%%%%%%%%%%%%%%%%%%%%%%%%%%% APPENDIX %%%%%%%%%%%%%%%%%%%%%%%%%%%%%%%%%%%%%%%
\appendix
\renewcommand{\thesection}{\Alph{section}.}
\renewcommand{\theequation}{\thesection \arabic{equation}}

\section{The case $\alpha=-1$}
\indent

The more general decomposition of the high momentum behavior of $T^{ij}({\bf k})$ for $\a=-1$ reads:
\begin{eqnarray}
\label{(A.1)}
T^{ij} ({\bf k}) &=& {\mathcal{B}}_{-1}\frac{\delta^{ij}}{|{\bf k}|}+\widetilde{{\mathcal{B}}}_{-1} \frac{{\bf k}^i{\bf k}^j}{|{\bf k} |^3}+ {\mathcal{P}}^{ij}({\bf k}) \, ,
\end{eqnarray}
where lim$_{{\bf k}\rightarrow \infty}{\mathcal{P}}^{ij}({\bf k}) \sim \frac{1}{{\bf k}^2}$.
Notice then that the integral in (\ref{(4.1)}) at most diverges logarithmically and, furthermore, the divergent term must be proportional to the $\delta^{ij}$ tensor.
By calculating the high energy behavior of the integral in the rhs of (\ref{(4.1)}) we obtain:
\begin{eqnarray}
\label{(A.2)}
 T^{ij} ({\bf k})&\sim & c_0 \delta^{ij} + c_1\frac{{\bf k}^i{\bf k}^j}{{\bf k}^2} -\frac{Mc_0}{2\pi^2}\left [ {{\mathcal{B}}}_{-1}+ \frac{\widetilde{{\mathcal{B}}}_{-1}}{3} \right ]{\rm{Log}}\left ( \frac{\; \Lambda^2}{-EM} \right )\delta^{ij} + \frac{Mc_1}{4\pi^2} \frac{{{\mathcal{B}}}_{-1}+\widetilde{{\mathcal{B}}}_{-1}}{3} \cdot \nonumber \\
&&  \cdot \left [ {\rm{Log}} \left ( \frac{{\bf k}^2}{\Lambda^2}\right )+ f_1 \right ]\delta^{ij}-\frac{Mc_1}{4\pi^2}\left [ {{\mathcal{B}}}_{-1}+ \frac{\widetilde{{\mathcal{B}}}_{-1}}{3} \right ] \left [ {\rm{Log}} \left ( \frac{\; {\bf k}^2}{-EM}\right )+ f_2 \right ] \frac{{\bf k}^i{\bf k}^j}{{\bf k}^2} \, ,
\end{eqnarray}    
with $f_1$ and $f_2$ two finite, constant terms.
Observe that, although the cut-off dependence can be removed by a suitable redefinition of $c_0$, the  non-analytic terms $\sim {\rm{Log}} \vert {\bf k}\vert$ cannot be compensated by the potential. Self-consistency of (\ref{(A.1)}) and (\ref{(A.2)}) force $c_1\rightarrow 0$ again.

\section{Proof of (\ref{(4.12)})}
\indent

Let us define:
\begin{eqnarray}
\label{(B.1)}
{\mathcal{H}}_{\alpha}(EM) &:= & c_1\int^{\Lambda} \frac{d^3k}{(2\pi)^3} \frac{T_{1\alpha}^{ii}({\bf k})}{E-{{\bf k}^2\over M}+i\eta} \quad \quad \alpha = 0, 1, 2  \, ,  
\end{eqnarray}
and concentrate on ${\mathcal{H}}_{0}(EM)$ (the analysis for ${\mathcal{H}}_{1}(EM)$ is identical). We have:
\begin{eqnarray}
\label{(B.2)}
{\mathcal{H}}_{0}(EM) = \int^{\Lambda} \frac{d^3k}{(2\pi)^3}\int^{\Lambda} \frac{d^3k^{\prime \prime}}{(2\pi)^3}\frac{c_1}{E-{{\bf k}^2\over M}+i\eta}\frac{({\bf k}-{\bf k^{\prime \prime}})^i ({\bf k}-{\bf k^{\prime \prime}})^j}{({\bf k}-{\bf k^{\prime \prime}})^2+m_{\pi}^2} \frac{T_0 ({\bf k^{\prime \prime}})}{E-{{\bf k}^2\over M}+i\eta} \nonumber \\
 +\int^{\Lambda} \frac{d^3k}{(2\pi)^3}\int^{\Lambda} \frac{d^3k^{\prime \prime}}{(2\pi)^3} \frac{c_1}{E-{{\bf k}^2\over M}+i\eta}\frac{c_1({\bf k}-{\bf k^{\prime \prime}})^i ({\bf k}-{\bf k^{\prime \prime}})^k+c_2\delta^{ik}}{({\bf k}-{\bf k^{\prime \prime}})^2+m_{\pi}^2} \frac{T_{10}^{kj}({\bf k^{\prime \prime}})}{E-{{\bf k}^2\over M}+i\eta} \, .
\end{eqnarray}

If we solve the equation above iteratively using (\ref{(4.4)}) for $T_0({\bf k})$ and (\ref{(4.7)}) for $T_{10}^{ij}({\bf k})$, the most divergent term in the $n$-th. iteration is (super-indexes $j$ and $k$ are contracted with unwritten momenta):
\begin{eqnarray}
\label{(B.3)}
&& c_1^{n+1}\left \{ \prod_{l=1}^{n+2}\int^{\Lambda} \frac{d^3 k_l}{(2\pi)^3} \right \}  \frac{1}{E-\frac{{\bf k}_1^2}{M}+i\eta} \frac{({\bf k}_1-{\bf k}_2)^i({\bf k}_1-{\bf k}_2)^j}{({\bf k}_1-{\bf k}_2)^2+m_{\pi}^2} \frac{1}{E-\frac{{\bf k}_2^2}{M}+i\eta} \, ... \nonumber \\
&& ... \,\frac{1}{E-\frac{{\bf k}_{n+1}^2}{M}+i\eta} \frac{({\bf k}_{n+1}-{\bf k}_{n+2})^k({\bf k}_{n+1}-{\bf k}_{n+2})^i}{({\bf k}_{n+1}-{\bf k}_{n+2})^2+m_{\pi}^2} \frac{1}{E-\frac{{\bf k}_{n+2}^2}{M}+i\eta} \, . 
\end{eqnarray}

Taking into account that the limits $E\rightarrow 0$ and $m_{\pi}^2\rightarrow 0$ exist and the flow (\ref{(4.9)}),
 the leading behavior in $\L$ reads: 
\begin{eqnarray}
\label{(B.4)}
(-M)^{n+2} c_1^{n+1} \left \{ \prod_{l=1}^{n+2}\int^{\Lambda} \frac{d^3 k_l}{(2\pi)^3} \frac{1}{{\bf k}_l^2} \right \}  \frac{({\bf k}_1-{\bf k}_2)^i({\bf k}_1-{\bf k}_2)^j}{({\bf k}_1-{\bf k}_2)^2} \, ... \, \frac{({\bf k}_{n+1}-{\bf k}_{n+2})^k({\bf k}_{n+1}-{\bf k}_{n+2})^i}{({\bf k}_{n+1}-{\bf k}_{n+2})^2} \sim   {\bar c}_1^{n+1} \L \, , \nonumber \\
\phantom{.}
\end{eqnarray}
which proves that $a_0$ is a (${\bar c}_1$-dependent) constant. Notice also 
that the integral in (\ref{(B.4)})
is bound by $(\int^{\L} d^3 {\bf k}/{\bf k}^2)^{n+2} $. Let us next identify the 
subleading behavior. Consider first 
 $E= 0$. The derivative of (\ref{(B.3)}) with respect to $m_{\pi}^2$ at
$m_{\pi}^2=0$ has at most a logarithmic singularity which means that the next to 
leading behavior in $\L$ is $\sim c_1^{n+1} \Lambda^{n-1}m_{\pi}^2{\rm{Log}} \L  $, which 
gives rise to ${\mathcal{O}}\left ( \frac{1}{\L} \right ) $ contributions in (\ref{(4.12)}). 
Terms contributing to $d_0$ in the $n$th. iteration appear when: (i) the $c_2$-proportional term of $T_0({\bf k})$ is iterated through only $c_1$ potential insertions coming from the second line in (\ref{(B.2)}); (ii) the equal to 1 term of $T_0({\bf k})$ is iterated in such a way that a $c_2$ potential from the last piece appears only once in the iteration. The relevant integral is obtained by substituting: 
\begin{eqnarray}
c_1\frac{({\bf k}_p-{\bf k}_{p+1})^i({\bf k}_p-{\bf k}_{p+1})^j}{({\bf k}_p-{\bf k}_{p+1})^2+m_{\pi}^2} \longrightarrow \frac{c_2 \, \delta^{ij}}{({\bf k}_p-{\bf k}_{p+1})^2+m_{\pi}^2} 
\end{eqnarray}
in (\ref{(B.3)}). In order to get the leading behavior in $\L$ of this integral we can set $m_{\pi}^2=0$ in all but the substituted term above. We have (super-indexes $j$, $l$, $q$ and $k$ are contracted with unwritten momenta):
\begin{eqnarray}
\label{(B.6)}
&&(-M)^{n+2} c_2 \, c_1^{n} \left \{ \prod_{l=1\setminus \{ p,p+1 \}}^{n} \int^{\Lambda} \frac{d^3 k_l}{(2\pi)^3} \frac{1}{{\bf k}_l^2} \right\} \frac{({\bf k}_1-{\bf k}_2)^i({\bf k}_1-{\bf k}_2)^j}{({\bf k}_1-{\bf k}_2)^2} \, ... \, \left [ \, \int^{\Lambda} \frac{d^3 k_p}{(2\pi)^3} \right. \nonumber \\
&& \left.  \int^{\Lambda} \frac{d^3 k_{p+1}}{(2\pi)^3} \frac{({\bf k}_{p-1}-{\bf k}_p)^l}{({\bf k}_{p-1}-{\bf k}_p)^2} \frac{1}{{\bf k}_p^2} \frac{({\bf k}_{p-1}-{\bf k}_p)\cdot({\bf k}_{p+1}-{\bf k}_{p+2})}{({\bf k}_p-{\bf k}_{p+1})^2+m_{\pi}^2}\frac{1}{{\bf k}_{p+1}^2} \frac{({\bf k}_{p+1}-{\bf k}_{p+2})^q}{({\bf k}_{p+1}-{\bf k}_{p+2})^2} \, \right ] \, ... \nonumber \\
&& ... \, \frac{({\bf k}_{n-1}-{\bf k}_n)^k({\bf k}_{n-1}-{\bf k}_n)^i}{({\bf k}_{n-1}-{\bf k}_n)^2} \sim  c_2 \, {\bar{c}}_1^{n} \,{\rm{Log}} \L \, ,
\end{eqnarray}
which proves (with the flow (\ref{(4.2)})) that $d_0$ is a constant.

Let us next address the energy dependent contribution to (\ref{(B.1)}). Notice that
any analytic contribution in $EM$ would show up at ${\mathcal{O}}(1/\L)$. Hence only non-analytic contributions (like the one in (\ref{(B.6)})) are relevant to us. Let us then look for non-analytic contributions in $EM$ in the most divergent diagram in the $n$th. iteration (\ref{(B.3)}). Since the $m_{\pi}^2\rightarrow 0$ limit exists we can take
 it and have:
\begin{eqnarray}
&&c_1^{n+1}\int^{\Lambda} \frac{dk_1}{(2\pi)^3}\frac{{\bf k}_1^2}{E-\frac{{\bf k}_1^2}{M}+i\eta} \, ... \, \int^{\Lambda}\frac{dk_{n+2}}{(2\pi)^3}\frac{{\bf k}_{n+2}^2}{E-\frac{{\bf k}_{n+2}^2}{M}+i\eta} \nonumber \\
&&\int d\Omega_1 \, ... \, \int d\Omega_{n+2} \frac{({\bf k}_1-{\bf k}_2)^i({\bf k}_1-{\bf k}_2)^j}{({\bf k}_1-{\bf k}_2)^2} \, ... \,  \frac{({\bf k}_{n+1}-{\bf k}_{n+2})^k({\bf k}_{n+1}-{\bf k}_{n+2})^i}{({\bf k}_{n+1}-{\bf k}_{n+2})^2} \, .
\end{eqnarray}
where $d\Omega_i$ , $i=1,...,n+2$ stand for angular integrals.
Since the most singular contribution comes from the region $\vert {\bf k}_{l}\vert \sim \L$ $\forall \,  l$, the angular integral will give rise to a constant (which, furthermore,  
is bound by $(4\pi )^{n+2}$), and the integrals over 
$\vert {\bf k}_{l} \vert$ decouple. Hence the leading behavior for small $E$ turns out to be the non-analytic contribution we are looking for ($\alpha_0$, $\beta_0$, ${\widetilde{\alpha}}_0$ and ${\widetilde{\beta}}_0$ are constants):
\begin{eqnarray}
\sim  c_1^{n+1}\left [ \int^{\Lambda}dk \frac{{\bf k}^2}{E-\frac{{\bf k}^2}{M}+i\eta} \right ]^{n+2}  &\sim &  c_1^{n+1}\left ( \alpha_0\Lambda + i\beta_0\sqrt{EM}+ {\mathcal{O}} \left ( \frac{1}{\Lambda} \right ) \right )^{n+2} \sim \nonumber \\
&& \sim \bar{c}_1^{n+1} \left ( \widetilde{\alpha }_0 \Lambda + i\widetilde{\beta }_0 \sqrt{EM} + {\mathcal{O}} \left ( \frac{1}{\Lambda} \right ) \right ) \, ,
\end{eqnarray}
which proves, in addition, that $b_0$ is a constant. Notice that a ${\rm{Log}} \L $ dependence in this term would have been fatal for renormalization.

We have then proved the first formula in (\ref{(4.12)}). The proof of the second formula is identical. The third formula is proved by simply noticing that all integrals involved are at most logarithmically divergent and, those which actually are, go multiplied by $c_1 \sim \frac{1}{\L}$.
           
\section{On $c_1$ tuning}
\indent

In section 4, when we focused on proving that a certain behavior of the bare constants of the potential as functions of the cut-off (namely, $c_0$, $c_1$ $\sim$ $\Lambda^{-1}$) would render a finite T-matrix, only $c_0$ was conveniently fine-tuned. As a result the so-computed scattering amplitude
lacked from partial wave mixing, which is expected due to the second rang tensorial term in the (bare) Hamiltonian. In order to obtain partial wave mixing 
two possibilities must be regarded. On 
the one hand, it could well happen that, indeed, mixing should not have been considered as LO, but as a NLO term to be treated perturbatively, the divergences it may cause being absorbed in the usual way by higher order local counterterms. This appears to be consistent with the fact that partial wave mixing in this channel amounts only to a few degrees. This treatment resums the $\delta^{ij}$-proportional part of OPE. Its SSB term, now eliminated by the strong suppression of $c_1$, is then recovered in a NLO analysis. We have shown how this works 
in the section 4.2 .

Nevertheless, another possibility remains unexamined. A proper tuning of $c_1$ to 
a, let's say, non-trivial 
RG fixed  point, could very well recover mixing at the leading order. So far, the existence of such a fixed point is anything but evident. Uncovering it or ruling it out requires detailed numerical work which is beyond the scope of this paper. However,
in order to illustrate our point let us provide two approximations that exemplify how this tuning would emerge, how it would affect previous results and to which extent to achieve this goal we depend on the exact resolution of our actual system of integral equations. 

Let's take in the following ${\bf k^{\prime}}={\bf 0}$ for simplicity. We will also apply the chiral limit ($m_{\pi}$, $c_2$ $\rightarrow $ 0) and work with
 ${\widetilde{c}}_0$ and ${\widetilde{c}}_1$ defined in section 4.2 . After decomposing the T-matrix in:
\begin{eqnarray}
T^{ij}({\bf k}) &=& T_1(k) \, \delta^{ij} +\left [ \frac{{\bf k}^i{\bf k}^j- \frac{{\bf k}^2}{3}\delta^{ij}}{{\bf k}^2} \right ] T_2(k) \, ,
\end{eqnarray} 
the following two angular integrals arise in the resolution of its LS equation: 
\begin{eqnarray}
{\widetilde{c}}_1 \int\frac{d\Omega^{\prime \prime}}{4\pi} & & \frac{({\bf k}-{\bf k^{\prime \prime}})^i ({\bf k}-{\bf k^{\prime \prime}})^j -\frac{({\bf k}-{\bf k^{\prime \prime}})^2}{3}\delta^{ij}}{({\bf k}-{\bf k^{\prime \prime}})^2} \,  \longrightarrow \, {\widetilde{c}}_1 \, \omega_1 \left ( \frac{k}{k^{\prime \prime}} \right ) \, \frac{ {\bf k}^i{\bf k}^j- \frac{{\bf k}^2}{3}\delta^{ij}}{{\bf k}^2} \, , \nonumber \\
{\widetilde{c}}_1 \int\frac{d\Omega^{\prime \prime}}{4\pi} & & \frac{({\bf k}-{\bf k^{\prime \prime}})^i ({\bf k}-{\bf k^{\prime \prime}})^k -\frac{({\bf k}-{\bf k^{\prime \prime}})^2}{3}\delta^{ik}}{({\bf k}-{\bf k^{\prime \prime}})^2}  \, \left [ \, \frac{{\bf k^{\prime \prime}}^k{\bf k^{\prime \prime}}^j - \frac{{\bf k^{\prime \prime}}^2}{3}\delta^{kj}}{{\bf k^{\prime \prime}}^2} \,  \right ]  \longrightarrow \nonumber \\
&& \longrightarrow  \left [ \,  {\widetilde{c}}_1 \, \omega_2 \left ( \frac{k}{k^{\prime \prime}} \right ) \, \frac{ {\bf k}^i{\bf k}^j- \frac{{\bf k}^2}{3}\delta^{ij}}{{\bf k}^2} + {\widetilde{c}}_1 \, \omega_3 \left ( \frac{k}{k^{\prime \prime}} \right ) \, \delta^{ij} \, \right ]  \, ,
\end{eqnarray} 
with $\omega_i\left (\frac{k}{k^{\prime \prime}} \right )$, i = 1, 2, 3, as known functions ($k = \vert {\bf k} \vert $, $k^{\prime \prime} = \vert {\bf k^{\prime \prime}} \vert$).

At this point we wish to introduce some reasonable approximation that allows us to transform the non-separable in $k$ and $k^{\prime \prime}$ functions $\omega_i \left ( \frac{k}{k^{\prime \prime}} \right )$ into separable ones. Once this is achieved we only need to solve a conventional system of equations and check whether, at least within this approximation, a non-trivial fixed point exists. Obviously our approximation should be as compatible as possible with what we know about the behavior of the full $d^3{\bf k}$-integrals. For instance:
\begin{eqnarray}  
\int^{\Lambda}\frac{dk^{\prime \prime}}{2\pi^2}\frac{k^{\prime \prime 2} \, \omega_i\left ( \frac{k}{k^{\prime \prime}} \right )}{E-\frac{k^{\prime \prime 2}}{M}+i\eta} \, \sim \, k \quad \quad \quad i=1,2 \, ,
\end{eqnarray}
that is, both are finite integrals proportional to $k$ in the limit $\Lambda \rightarrow \infty $. Unfortunately, no separable $\omega_i$ achieves this. We shall content ourselves with a simple but still reasonable starting point that enforces separability.
Then, let us take $\omega_3 \left ( \frac{k}{k^{\prime \prime}} \right )$ as a constant ($:= \alpha_3$) and substitute $\omega_{1, 2} \left ( \frac{k}{k^{\prime \prime}} \right )$ by $:= \alpha_{1,2} \, \frac{k}{k^{\prime \prime}} $ ($\alpha_{1, 2}$ also being constants). Although the latter introduces logarithmic divergences which do not exist in the actual function, it keeps the correct behavior in $k$ shown in (C.3).

The LS equation takes the form:
\begin{eqnarray}
\label{(C.4)}
T^{ij}({\bf k}) &=& {\widetilde{c}}_0(1+T_1 \,{\mathcal{I}}_0) \, \delta^{ij}+{\widetilde{c}}_1\left [ \frac{{\bf k}^i{\bf k}^j- \frac{{\bf k}^2}{3}\delta^{ij}}{{\bf k}^2} \right ]+ {\widetilde{c}}_1\alpha_1 \left [ \frac{{\bf k}^i{\bf k}^j- \frac{{\bf k}^2}{3}\delta^{ij}}{{\bf k}^2} \right ] \, k \, T_1 \, {\mathcal{I}}_{-1} + \nonumber \\
&& + {\widetilde{c}}_1\alpha_2 \left [ \frac{{\bf k}^i{\bf k}^j- \frac{{\bf k}^2}{3}\delta^{ij}}{{\bf k}^2} \right ] \, k \,  {\mathcal{C}}+ {\widetilde{c}}_1 \, \alpha_3 \,  \delta^{ij}  \, {\mathcal{B}}\, ,
\end{eqnarray}
where we have already used that $T_1(k)$ becomes momentum independent, as it is easily verified through (\ref{(C.4)}):
\begin{eqnarray}
T_1 &=& {\widetilde{c}}_0(1+T_1 \, {\mathcal{I}}_0) +{\widetilde{c}}_1 \, \alpha_3 \, {\mathcal{B}} \, ,  \nonumber \\
T_2 (k) &=& {\widetilde{c}}_1(1+\alpha_1\, k \,  T_1\, {\mathcal{I}}_{-1}+\alpha_2 \, k \, {\mathcal{C}} ) \, .
\end{eqnarray}

The following functions have been introduced:
\begin{eqnarray}
{\mathcal{I}_0} &:=& \int^{\Lambda} \frac{d k}{2\pi^2} \frac{k^2}{E-\frac{k^2}{M}+i\eta}  \, , \nonumber \\
{\mathcal{B}} &:=&  \int^{\Lambda}\frac{d k}{2\pi^2} \frac{k^2 \, T_2(k)}{E-\frac{k^2}{M}+i\eta} \, , \nonumber \\
{\mathcal{I}}_{-1} &=& \int^{\Lambda} \frac{d k}{2\pi^2} \frac{k}{E-\frac{k^2}{M}+i\eta} \, , \nonumber \\
{\mathcal{C}} &=& \int^{\Lambda} \frac{d k}{2\pi^2} \frac{k \, T_2(k)}{E-\frac{k^2}{M}+i\eta} \, .
\end{eqnarray}
 
A few manipulations allow us to solve for $T_1$ and the combination $\alpha_1 \, T_1 \, {\mathcal{I}}_{-1}+\alpha_2 \, {\mathcal{C}}$:
\begin{eqnarray}
\label{(C.7)}
T_1 &=& \frac{{\widetilde{c}}_0+{\widetilde{c}}_1^2\alpha_3{\mathcal{I}}_0+{\widetilde{c}}_1^3 \alpha_1\alpha_3 \frac{{\mathcal{I}}_1{\mathcal{I}}_{-1}}{1-{\widetilde{c}}_1\alpha_2{\mathcal{I}}_0}}{1-{\widetilde{c}}_0{\mathcal{I}}_0 -{\widetilde{c}}_1^2\alpha_1\alpha_3 \frac{ {\mathcal{I}}_1{\mathcal{I}}_{-1}}{1-{\widetilde{c}}_1\alpha_2{\mathcal{I}}_0}} \, , \nonumber \\
\alpha_1 \, T_1{\mathcal{I}}_{-1}+\alpha_2 \, {\mathcal{C}} &=& \left ( \frac{{\widetilde{c}}_1\alpha_2 {\mathcal{I}}_{-1}}{1-{\widetilde{c}}_1\alpha_2 {\mathcal{I}}_0} \right ) \frac{\left ( {\widetilde{c}}_0+{\widetilde{c}}_1^2\alpha_3{\mathcal{I}}_0 \right ) \frac{\alpha_1}{\alpha_2}+{\widetilde{c}}_1 (1-{\widetilde{c}}_0{\mathcal{I}}_0)}{1-{\widetilde{c}}_0{\mathcal{I}}_0-{\widetilde{c}}_1^2\alpha_1\alpha_3 \frac{ {\mathcal{I}}_1{\mathcal{I}}_{-1}}{1-{\widetilde{c}}_1\alpha_2{\mathcal{I}}_0}} \,
\end{eqnarray}
where a quadratic divergence:
\begin{eqnarray}
{\mathcal{I}}_1:= \int^{\Lambda}\frac{d k}{2\pi^2} \frac{ k^3}{E-\frac{k^2}{M}+i\eta} \, ,
\end{eqnarray}
enters. 

It is not difficult to realize that little has been gained:
the only way to get (\ref{(C.7)}) finite is by an untuned ${\widetilde{c}}_1$, (${1-{\widetilde{c}}_1\alpha_2{\mathcal{I}}_0}\not=0 $), and a tuned ${\widetilde{c}}_0$, which force $T_2(k)$ to become trivial again. We have not been able to figure out any
reasonable approximation which produces a non-trivial $T_2(k)$. 
  
Anyway, in order to illustrate the kind of fixed point we are looking for,
let us take another option which, unfortunately, is completely unrealistic. It
 consists of sending $\omega_1 \left ( \frac{k}{k^{\prime \prime}} \right )$ and $\omega_3 \left ( \frac{k}{k^{\prime \prime}} \right )$ to zero, keeping $\omega_2 \left ( \frac{k}{k^{\prime \prime}} \right )$ as a mere constant ($:= \alpha_2$). That presents the main advantage of producing decoupled equations for $T_1$ and $T_2$:
\begin{eqnarray}
T_1 &=& {\widetilde{c}}_0(1+{\mathcal{A}}) \, , \nonumber \\
T_2 &=& {\widetilde{c}}_1(1+\alpha_2 \, {\mathcal{B}}) \, .
\end{eqnarray} 
where ${\mathcal{A}}:= T_1 \, {\mathcal{I}}_0$ and ${\mathcal{B}} = T_2 \, {\mathcal{I}}_0$.  
Both are well defined provided ${\widetilde{c}}_0(1+{\mathcal{A}})$ and ${\widetilde{c}}_1(1+\alpha_2 \, {\mathcal{B}})$ are finite. We compute them multiplying above by $1/(E-\frac{{\bf k}^2}{M}+i\eta)$ and integrating. This produces:
\begin{eqnarray}
{\widetilde{c}}_0(1+{\mathcal{A}}) &=& \frac{1}{\frac{1}{{\widetilde{c}}_0}-{\mathcal{I}}_0} \, , \nonumber \\
{\widetilde{c}}_1(1+\alpha_2 \, {\mathcal{B}}) &=& \frac{1}{\frac{1}{{\widetilde{c}}_1}-\alpha_2 \, {\mathcal{I}}_0} \, .
\end{eqnarray} 

It is obvious that divergences are absorbed if ${\widetilde{c}}_0$, ${\widetilde{c}}_1$ behave like $\Lambda^{-1}$ and non-trivial results ($T_1$, $T_2$ $\ne$ 0) require:
\begin{eqnarray}
\frac{1}{{\widetilde{c}}_0} &:=& -\frac{M\Lambda }{2\pi^2}+ \frac{1}{{\widetilde{c}}_0^r(\mu )} \, , \nonumber \\
\frac{1}{{\widetilde{c}}_1} &:=& -\frac{M\Lambda \alpha_2 }{2\pi^2} + \frac{1}{{\widetilde{c}}_1^r(\mu )} \, .
\end{eqnarray}

Namely, ${\widetilde{c}}_1$ must be fine-tuned (to a non-trivial fixed point)
as desired. Unfortunately, as mentioned before, the assumptions made for the $\omega_i$ here are not 
realistic.

Summarizing, we are rather pessimistic about the possibility that a non-trivial RG fixed point for both ${\widetilde{c}}_0$ and ${\widetilde{c}}_1$ exists,
which allows for partial wave mixing at leading order.

\section{Proof that no continuous solution of eq. (18) of \cite{Bedaque}
exists when $R\rightarrow 0$}
\indent

Consider eq. (18) of \cite{Bedaque},
\begin{eqnarray}
\label{disc}
\sqrt{-MV_0}R \cot (\sqrt{-MV_0}R )={3\over 4} + \sqrt{6M\alpha_{\pi}\over R}
\tan (2\sqrt{6M\alpha_{\pi}\over R}+\phi_0 ) \, ,
\end{eqnarray}
with $V_0 < 0$ , $R > 0$, $\phi_0\in [-\pi /2 , \pi /2 ]$. We are interested in whether continuous solutions $V_0=V_0 (R)$ exist when $R\rightarrow 0$. Let us define:
\begin{eqnarray}
y:=\sqrt{-MV_0}R > 0 \quad ,\quad x:=2\sqrt{6M\alpha_{\pi}\over R}+\phi_0 \, .
\end{eqnarray}
In terms of these variables we are interested in whether a continuous solution  $y=y(x)$ exists when $x\rightarrow \infty$ for the following equation,
\begin{eqnarray}
y\cot y = {3\over 4} + {x-\phi_0\over 2} \tan x \, .
\end{eqnarray}
Deriving this equation once one obtains:
\begin{eqnarray}
\label{multbr}
{\sin 2y -2y \over 2 \sin^2 y}{d y\over d x}={\sin 2x +2x -2\phi_0\over 4 \cos^2 x} \, ,
\end{eqnarray}
which proves that $y(x)$ decreases when $x$ increases for $x$ large enough. The proof holds everywhere except for the points $x=(n+1/2)\pi $, $y=m\pi$, $n,m=0,1,2...$ which we analyze in the following.

When $x$ approaches $(n+1/2)\pi$ for a given $n$, $y$ must necessarily approach $m\pi$ for some $m$ in order for eq. (\ref{multbr}) to have a solution. If we write:
\begin{eqnarray}
x=(n+{1\over 2})\pi + \delta x \quad ,\quad y=m\pi +\delta y \quad, \quad
\delta x \; ,\; \delta y \rightarrow 0 \, ,
\end{eqnarray}
we have, for $m\not= 0$,
\begin{eqnarray}
\delta y = -{2m\over n +{1\over 2}-{\phi_0\over \pi}} \delta x + {\mathcal O}({\delta x}^2) 
\end{eqnarray}
Hence, eq. (\ref{multbr}) admits a continuous solution near the point $x=(n+1/2)\pi$
provided that we choose $m\not=0$. Notice also that $y$ keeps decreasing when
$x$ increases in the neighborhood of this point. 

Now, if we increase $x$ from $(n+1/2)\pi$ to $(n+3/2)\pi$, $y$ must decrease
from $m\pi$ to $(m-1)\pi$, if continuity is required. By iterating the argument, if we increase $x$ till $(n+m+1/2)\pi$, continuity requires $y$ to decrease till $0$.
However for $x= (n+m+1/2)\pi+\delta x$ ($\delta x \rightarrow 0$) and $y=\delta y\rightarrow 0$, eq. (\ref{multbr}) does not have a solution anymore, since one obtains:
\begin{eqnarray}
1+{\mathcal O}({\delta y}^2) = -{(n+m+{1\over 2})\pi -\phi_0\over 2\delta x} + {\mathcal O}(1)
\end{eqnarray}
%which has no solution for $\delta x \rightarrow 0$.
This implies, in particular, that the curves plotted in fig. 4 of \cite{Bedaque} cannot be continuously extended below $R\sim 0.25 \, fm.$, $R\sim 0.13\, fm.$ and $R\sim 0.09 \, fm.$ respectively, as it shown in the figure below. 

\begin{figure}
\begin{center}
\begin{boxedminipage}[t]{10cm}
\vspace{5mm}
\makebox[0truecm]{\phantom a}
{\epsfxsize=8.0truecm\epsffile{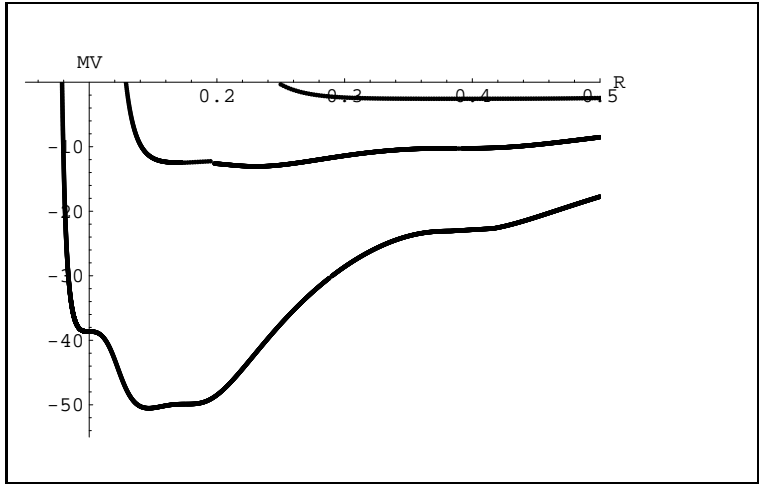}}
\vspace{5mm}
\end{boxedminipage}

\caption{\small In the figure it is shown how the first three branches of the flow presented in Figure 4 of \cite{Bedaque}, behave as one approaches the relevant limit $R \rightarrow $ 0.} 

\end{center}
\end{figure}

In conclusion, no continuous solution $y=y(x)$ of eq. (\ref{multbr}) (and hence of eq. (\ref{disc})) exists for $x\rightarrow \infty$ ($R\rightarrow 0$). If continuity is given up, an infinite number of solutions exist, none of them being compatible with a RG flow, at least in the standard sense. Note also that this situation is qualitatively different from a limit cycle behavior \cite{Wilson}, which is realized, for instance, in three body systems \cite{3body}. There the flows are oscillating and discontinuous but uniquely defined, no matter how large the cut-off is.

\newpage

%%%%%%%%%%%%%%%%%%%%%%%%%%%%% BIBLIOGRAPHY %%%%%%%%%%%%%%%%%%%%%%%%%%%%%%%%%%%%%


\begin{thebibliography}{99}

\bibitem{Weinberg} S. Weinberg, Phys. Lett. {\bf B251}, 288 (1990); Nucl. Phys. {\bf B363}, 3 (1991); Phys. Lett. {\bf B295}, 114 (1992).

\bibitem{reviews} U. van Kolck, Prog. Part. Nucl. Phys. {\bf 43}, 337 (1999); S. R. Beane et al., {\sl nucl-th/0008064}.

\bibitem{JM} E. Jenkins and A. V. Manohar, Phys. Lett. {\bf B255}, 558 (1991).

\bibitem{VK} C. Ordo\~{n}ez, L. Ray and U. van Kolck, Phys. Rev. {\bf C53}, 2086 (1996).

\bibitem{Meissner} E. Epelbaum, W.Gl$\ddot{\mathrm{o}}$ckle, Ulf-G. Mei{\selectlanguage{german}"s}ner, Nucl. Phys. {\bf A671}, 295 (2000); Nucl.Phys. {\bf A637}, 107 (1998).

\bibitem{Kaiser} N. Kaiser, Phys. Rev. {\bf C61}, 014003 (2000); Phys. Rev. {\bf C62}, 024001 (2000); Phys. Rev. {\bf C63}, 044010 (2001).

\bibitem{Walzl} M. Walzl, Ulf-G. Mei{\selectlanguage{german}"s}ner and E. Epelbaum, Nucl. Phys. {\bf A693} 663 (2001).
 
\bibitem{Mont} A. Pineda and J. Soto, Nucl. Phys. (Proc. Suppl.) {\bf B64}, 428 (1998).
 
\bibitem{Bern3} J. Gasser, V. E. Lyubovitskij, A. Rusetsky and A. Gall, Phys. Rev. {\bf D64} 016008 (2001).

\bibitem{Pionium} D. Eiras and J. Soto, Phys. Rev. {\bf D61}, 114027 (2000); $\pi$N Newslett. {\bf 15}, 181 (1999).

\bibitem{KSW} D. B. Kaplan, M. J. Savage and M. B. Wise, Phys. Lett. {\bf B424}, 390 (1998).

\bibitem{Stewart} S. Fleming, T. Mehen and I. W. Stewart, Nucl. Phys. {\bf A677}, 313 (2000); Phys. Rev. {\bf C61}, 044005 (2000).

\bibitem{Phillips} D. R. Phillips, {\sl nucl-th/9804040}.

\bibitem{Beane} D. R. Phillips, S. R. Beane and T. D. Cohen, Annals Phys. {\bf 263}, 255 (1998).

\bibitem{Steele} J. V. Steele and R. J. Furnstahl, Nucl. Phys. {\bf A637}, 46 (1998).

\bibitem{StewartMehen} T. Mehen and I. W. Stewart, Phys. Lett. {\bf B445}, 378 (1999); J. Gegelia, {\sl nucl-th/9806028}.

\bibitem{Kolck} U. van Kolck, Prog. Part. Nucl. Phys. {\bf 43}, 337 (1999); Nucl. Phys. {\bf A645}, 273 (1999).

\bibitem{Kaplan} D. B. Kaplan, M. J. Savage and M. B. Wise, Nucl. Phys. {\bf B478}, 629 (1996).

\bibitem{Bedaque} S. R. Beane, P. F. Bedaque, M. J. Savage and U. van Kolck, Nucl. Phys. {\bf A700}, 377 (2002).

\bibitem{Frederico} T. Frederico, V. S. Tim{\'o}teo and L. Tomio, Nucl. Phys. {\bf A653}, 209 (1999).

\bibitem{LaPlata} H. E. Camblong et al., Phys. Rev. Lett. {\bf 85}, 1590 (2000); Annals Phys. {\bf 287}, 14 and 57 (2001).

\bibitem{LL} N. Brambilla, A. Pineda, J. Soto and A. Vairo, Phys. Lett. {\bf B470}, 215 (1999).

\bibitem{short} N. Brambilla, A. Pineda, J. Soto and A. Vairo, Phys. Rev. {\bf D60}, 091502 (1999); Nucl. Phys. {\bf B566}, 275 (2000).

\bibitem{Lepage} P. Lepage, {\sl nucl-th/9706029}.

\bibitem{Kaiserhw} N. Kaiser, S. Gerstendorfer, W. Weise, 
    Nucl. Phys. {\bf A637}, 395 (1998).

\bibitem{Childress} S. R. Beane et al., Phys. Rev. {\bf A64} 042103 (2001); D. W. L. Sprung et al., Phys. Rev. {\bf C49}, 2942 (1994).

\bibitem{Lepage2} G.P. Lepage, {\it What is renormalization?}, Boulder TASI 1989:483-508  (QCD161:T45:1989).
 
\bibitem{Wilson} S. D. Glazek, K. G. Wilson, Phys. Rev. {\bf D47}, 4657 (1993);
hep-th/0203088.

\bibitem{3body} P.F. Bedaque, H.W. Hammer, U. van
Kolck, Nucl. Phys. {\bf A646}, 444 (1999); Nucl. Phys. {\bf A676}, 357 (2000); P.F. Bedaque, E. Braaten, 
H.W. Hammer, Phys. Rev. Lett. {\bf 87}, 160407 (2001); H.W. Hammer, T. Mehen, Phys. Lett. {\bf B516}, 353 (2001).
\end{thebibliography}
\end{document}